\documentclass[10pt,journal]{IEEEtran}

\usepackage[dvipsnames]{xcolor}
\usepackage[normalem]{ulem}

\newcommand*{\tran}{\mathsf{T}}
\newcommand*{\herm}{\mathsf{H}}

\usepackage{cite}

\usepackage{graphicx}
\graphicspath{{./figures/}}
\DeclareGraphicsExtensions{.png,.jpg,.jpeg,.pdf}

\usepackage{amsmath,amssymb,amsthm,amsfonts,amscd,amsbsy,amsxtra,amsfonts}
\usepackage{mathtools}

\usepackage{booktabs}
\usepackage{multirow}

\usepackage[caption=false,font=footnotesize]{subfig}

\DeclareMathOperator{\diag}{diag}

\DeclareMathOperator*{\argmin}{arg\,min}
\DeclareMathOperator{\var}{Var}
\DeclareMathOperator{\cov}{Cov}
\DeclareMathOperator{\mse}{MSE}

\DeclarePairedDelimiter\abs{\lvert}{\rvert}%
\DeclarePairedDelimiter\norm{\lVert}{\rVert}%

\newtheorem{remark}{\underline{Remark}}
\newtheorem{proposition}{\underline{Proposition}}
\newtheorem{definition}{\underline{Definition}}

\begin{document}
%
\title{Performance of Orthogonal Delay-Doppler Division Multiplexing Modulation with Imperfect Channel Estimation}

\author{
    Kehan~Huang,~\IEEEmembership{Student~Member,~IEEE,}
    Min~Qiu,~\IEEEmembership{Member,~IEEE,}
    Jun~Tong,~\IEEEmembership{Member,~IEEE,}
    Jinhong~Yuan,~\IEEEmembership{Fellow,~IEEE,}
    and~Hai~Lin,~\IEEEmembership{Senior~Member,~IEEE}

    \thanks{The work of K. Huang, M. Qiu, and J. Yuan was supported in part by the Australian Research Council (ARC) Discovery Project under Grant DP220103596, and in part by the ARC Linkage Project under Grant LP200301482. The work of Hai Lin was supported by the Japan Society for the Promotion of Science (JSPS) Grants-in-Aid for Scientific Research (KAKENHI) under Grant 22H01491. This work has been presented in part at the 2023 IEEE 24th International Workshop on Signal Processing Advances in Wireless Communications \cite{Huang2023ODDM_Performance}.}
    \thanks{K. Huang, M. Qiu, and J. Yuan are with the School of Electrical Engineering and Telecommunications, University of New South Wales, Sydney, NSW 2052, Australia, (email: kehan.huang@unsw.edu.au; min.qiu@unsw.edu.au; j.yuan@unsw.edu.au).}
    \thanks{J. Tong is with the School of Electrical, Computer and Telecommunications Engineering, University of Wollongong, Wollongong, NSW 2522, Australia (e-mail: jtong@uow.edu.au).}
    \thanks{H. Lin is with the Department of Electrical and Electronic Systems Engineering, Graduate School of Engineering, Osaka Metropolitan University, Sakai, Osaka 599-8531, Japan (e-mail: hai.lin@ieee.org).
    }
}

\maketitle

\begin{abstract}
    The orthogonal delay-Doppler division multiplexing (ODDM) modulation is a recently proposed multi-carrier modulation that features a realizable pulse orthogonal with respect to the delay-Doppler (DD) plane's fine resolutions. In this paper, we investigate the performance of ODDM systems with imperfect channel estimation considering three detectors, namely the message passing algorithm (MPA) detector, iterative maximum-ratio combining (MRC) detector, and successive interference cancellation with minimum mean square error (SIC-MMSE) detector. We derive the post-equalization signal-to-interference-plus-noise ratio (SINR) for MRC and SIC-MMSE and analyze their bit error rate (BER) performance. Based on this analysis, we propose the MRC with subtractive dither (MRC-SD) and soft SIC-MMSE initialized MRC (SSMI-MRC) detector to improve the BER of iterative MRC. Our results demonstrate that soft SIC-MMSE consistently outperforms the other detectors in BER performance under perfect and imperfect CSI. While MRC exhibits a BER floor above $10^{-5}$, MRC-SD effectively lowers the BER with a negligible increase in detection complexity. SSMI-MRC achieves better BER than hard SIC-MMSE with the same detection complexity order. Additionally, we show that MPA has an error floor and is sensitive to imperfect CSI.
\end{abstract}


\section{Introduction}\label{sec:introduction}

The widely adopted orthogonal frequency-division multiplexing (OFDM) technology modulates information symbols in the time-frequency (TF) domain. Although OFDM is good at combating frequency selective fading in linear time-invariant (LTI) channels, it is vulnerable to inter-carrier interference (ICI) in high-mobility environments, such as aircraft and satellite communications. As wireless communications are moving to higher frequency bands, ICI becomes particularly relevant due to the more prominent Doppler effect. This motivates us to consider the doubly-selective channel model, where both the user mobility and distance significantly impact the signal propagation.

In this context, orthogonal time frequency space (OTFS) modulation was proposed \cite{Hadani2017OrthogonalModulation, Hadani2018OTFS:5G} to modulate information symbols in the delay-Doppler (DD) domain, which facilitates the design of transmit signals to effectively couple with the channel's DD spreading functions.

Ideally, OTFS modulation requires its pulses to satisfy the biorthogonal robust condition. However, such pulses do not exist due to the uncertainty principle \cite{Lin2022OrthogonalModulation, Raviteja2018InterferenceModulation}. Instead, the rectangular pulse-shaped OTFS is usually considered in the literature \cite{Tiwari2019LowOTFS, Thaj2020LowSystems, Mishra2021IterativePilots, Zhang2021ASystem, Thaj2022GeneralVariants, Li2022CrossModulation, Li2022IterativeModulation}. As pointed out in \cite{Lin2022OrthogonalModulation}, the use of rectangular pulses introduces high out-of-band emission (OOBE). In addition, when a practically necessary bandpass filter is employed at the receiver, rectangular pulses contribute to complicated intersymbol interference (ISI) in OTFS systems \cite{Lin2022OrthogonalModulation, Shen2022ErrorReceivers}. To address these issues, the orthogonal delay-Doppler division multiplexing (ODDM) modulation is proposed with the delay-Doppler plane orthogonal pulse (DDOP) \cite{Lin2022OrthogonalModulation,Lin2022OnPulse,Lin2023TF2DD}. DDOP uses a square-root Nyquist pulse train to achieve sufficient orthogonality on the DD plane with fine resolutions \cite{Lin2022OnPulse,Lin2023TF2DD}. By selecting an appropriate square-root Nyquist pulse, such as the root raised cosine pulse, low OOBE can be achieved for ODDM. Furthermore, without the complicated ISI from rectangular pulses, ODDM has an \emph{exact} DD domain channel input-output relation, which can be exploited by the receiver to accurately detect the transmitted signals.

Because the doubly-selective channel spreads the signal across the DD domain, each received OTFS/ODDM symbol is superimposed by components from multiple symbols with different delay and Doppler indices. This enhanced path resolvability in the DD domain increases the available channel diversity, allowing the receiver to achieve better error performance. However, unlike a LTI channel, which can be divided into multiple frequency-flat subchannels corresponding to a single-tap equalizer, the doubly-selective channel introduces more complicated ISI and ICI, significantly increasing detection complexity. This motivates the need for designing low-complexity detectors. Although numerous detectors have been proposed for OTFS \cite{Tiwari2019LowOTFS, Raviteja2018InterferenceModulation, Thaj2020LowSystems, Zhang2021ASystem, Li2022IterativeModulation, Thaj2022GeneralVariants, Li2022CrossModulation}, a comprehensive performance comparison of these detectors for the recently developed ODDM remains lacking. Moreover, few studies have analyzed the performance of the detection algorithms under imperfect receiver channel state information (CSI). To the best of our knowledge, \cite{Naikoti2022PerformanceInformation} is the only study that has analyzed the impact of channel estimation errors in the context of OTFS by considering a maximum-likelihood (ML) detector, which has a prohibitive complexity for practical applications.

In this paper, we investigate the performance of ODDM over high-mobility channels with imperfect channel estimation. We first consider a channel estimation scheme based on embedded pilots. Three representative detectors are selected for comparison: the message passing algorithm (MPA) detector, the iterative maximum-ratio combining (MRC) detector, and the successive interference cancellation with minimum mean square error (SIC-MMSE) detector. For SIC-MMSE detectors, both hard and soft interference cancellation are considered. Note that the original MRC and SIC-MMSE detectors were designed for zero-padded (ZP)-OTFS \cite{Thaj2020LowSystems,Li2024SICMMSE_Turbo}, in order to reduce ISI and detection complexity. We study the performance of these detection algorithms for ODDM by catering for its distinct input-output relation with a frame-wise cyclic prefix (CP). For the first time, we analyze the signal-to-interference-plus-noise ratio (SINR) and bit error rate (BER) of MRC and SIC-MMSE for ODDM with imperfect CSI, and comprehensively compare their performance with MPA. Our detailed contributions are as follows.

\begin{itemize}
    \item To analyze the impact of imperfect CSI on iterative MRC and SIC-MMSE, a closed-form expression is derived for their post-equalization SINR. In particular, we consider channel estimation errors motivated by an embedded pilot-aided channel estimation scheme. Our derived SINR closely aligns with the simulation results. Through SINR analysis, we prove the equivalence of MRC and SIC-MMSE from the second iteration onwards. Furthermore, a tight SINR upper bound is derived to demonstrate that both iterative MRC and SIC-MMSE can effectively reduce interference power under imperfect CSI.
    
    \item Based on the derived SINR, we use state evolution to analyze the BER of MRC and SIC-MMSE under Gaussian residual interference. The theoretical analysis provides good approximations of the actual BER performance for both hard and soft SIC-MMSE detectors. We also reveal the limitation of MRC in fully canceling non-Gaussian residual interference, which causes an error floor in the high signal-to-noise ratio (SNR) region.

    \item Based on our BER analysis, we propose two methods to improve the BER of iterative MRC. Firstly, we design the MRC with subtractive dither (MRC-SD) detector to effectively lower the error floor of MRC with a negligible increase in computational complexity. We prove the decorrelation effect of a uniformly distributed dither signal for symbol estimation errors within a bounded region. Then, we propose a soft SIC-MMSE initialized MRC (SSMI-MRC) detector to approach the BER performance of soft SIC-MMSE while getting the same computational complexity as that of hard SIC-MMSE.

    \item Finally, we provide a comprehensive performance comparison of MPA, MRC, and SIC-MMSE for the ODDM system. We evaluate their BER, convergence characteristics, robustness against imperfect CSI, and complexity tradeoffs. Our results demonstrate that SIC-MMSE achieves the best error performance among the three detectors. In addition, both SIC-MMSE and MRC show faster convergence speed and higher robustness to imperfect CSI than MPA.
\end{itemize}

\textbf{Notations}: $(.)^*$ and $(.)^\herm$ denote the complex conjugate and Hermitian transpose, respectively. $\mathbf{F}_N$ denotes the normalized $N$-point discrete Fourier transform (DFT) matrix. $[.]_N$ means modulo $N$. $|a|$ outputs the absolute value of complex number $a$. $\norm{\mathbf{A}}$ outputs the induced 2-norm of matrix $\mathbf{A}$. Denote the vectorization of $\mathbf{A}\in\mathbb{C}^{m\times n}$ by $\mathbf{a}=\mathrm{vec}(\mathbf{\mathbf{A}})$, and denote its inverse operation by $\mathrm{vec}_{m\times n}^{-1}(\mathbf{a})=\mathbf{\mathbf{A}}$. $|\mathcal{A}|$ outputs the cardinality of set $\mathcal{A}$. $\mathbb{Z}[i]$ denotes the set of Gaussian integers.

\section{ODDM System Model}\label{sec:oddm_sys}

In this section, we introduce the ODDM system model proposed in \cite{Lin2022OrthogonalModulation} and investigated in \cite{Tong2024ODDM_PhyChan}. We define variables and a sub-input-output relation to concisely illustrate MRC and SIC-MMSE. This model serves as the foundation necessary for our subsequent analysis.

\subsection{DD Domain Communications}

Consider a DD grid with a time resolution of $\mathcal{T}_\mathrm{DD}=\frac{T}{M}$ and a frequency resolution of $\mathcal{F}_\mathrm{DD}=\frac{1}{NT}$. To arrange $MN$ information-bearing symbols onto the DD grid, a delay-Doppler multi-carrier (DDMC) signal transmits $M$ multi-carrier symbols, each staggered by $\frac{T}{M}$ and consisting of $N$ subcarriers spaced by $\frac{1}{NT}$. Let $\mathbf{X}\in\mathbb{C}^{M\times N}$ denote the DD domain symbols. The continuous-time DDMC signal is
\begin{equation}\label{tx}
    s(t)=\sum_{m=0}^{M-1}\sum_{n=-\frac{N}{2}}^{\frac{N}{2}-1} X[m,n] g_{tx}\left(t-m\frac{T}{M}\right)e^{j2\pi\frac{n}{NT}(t-m\frac{T}{M})}, \nonumber
\end{equation}
where $X[m,n]$ denotes the symbol in $\mathbf{X}$ at the $m$-th delay and $n$-th Doppler and $g_{tx}(t)$ is the transmit pulse.

Suppose the signal goes through a doubly-selective channel with $P$ resolvable paths. The channel has a deterministic representation in the DD domain as  $h(\tau,\nu)=\sum_{p=1}^{P} h_p \delta (\tau-\tau_p)\delta (\nu-\nu_p)$, where $\delta(\cdot)$ denotes the Dirac delta function, with $h_p$, $\tau_p$, and $\nu_p$ being the channel gain, delay shift, and Doppler shift of the $p$-th path, respectively \cite{Bello1963CharacterizationChannels}. Then, the received signal is
\begin{equation}
    r(t) = \int\int h(\tau,\nu)s(t-\tau)e^{j2\pi\nu(t-\tau)}d\tau d\nu + z(t),
\end{equation}
where $z(t)\sim\mathcal{CN}(0,\sigma_z^2)$ is the additive white Gaussian noise (AWGN). At the receiver, $r(t)$ is matched filtered by a receive pulse $g_{rx}(t)$ and converted back to the DD domain:
\begin{equation}\label{rx}
    Y[m,n] = \int r(t) g_{rx}^*\left(t-m\frac{T}{M}\right)e^{-j2\pi\frac{n}{NT}(t-m\frac{T}{M})}dt.
\end{equation}

\subsection{ODDM Transmitter}\label{sec:oddm_tx}
To fulfill the bi-orthogonality requirement with respect to the DD resolution, ODDM deploys the DDOP $u(t)$ as the practical waveform. Specifically, $g_{tx}(t)$ and $g_{rx}(t)$ are chosen as the DDOP given by
\begin{equation}
    u(t) = \sum_{\dot{n}=0}^{N-1}a(t-\dot{n}T),
\end{equation}
with $a(t)$ being a truncated square-root Nyquist pulse spanning $S$ symbol intervals on each side, i.e., $T_a=2S\frac{T}{M}$. By using a square-root Nyquist subpulse train, $u(t)$ satisfies the bi-orthogonality condition within the region $\abs{m}\leq M-1$ and $\abs{n}\leq N-1$ \cite{Lin2022OrthogonalModulation}. It should be noted that the existence of DDOP does not violate the uncertainty principle as it provides sufficient bi-orthogonality rather than global bi-orthogonality \cite{Lin2022OnPulse,Lin2023TF2DD}.

In practical implementation, a more efficient IDFT-based method is used to closely approximate the ODDM waveform when $2S \ll M$. $\mathbf{X}$ is firstly converted into the delay-time (DT) domain by $N$-point IDFT: $(\mathbf{X}^{\mathrm{DT}})^\tran = \mathbf{F}_N^\herm \mathbf{X}^\tran$. We then vectorize $\mathbf{X}^{\mathrm{DT}}$ to obtain the time domain digital samples, written as $\mathbf{s} = \mathrm{vec}(\mathbf{X}^{\mathrm{DT}})$. Finally, the approximated ODDM waveform is generated by $a(t)$-based filtering \cite{Lin2022OrthogonalModulation,Lin2023TF2DD}:
\begin{align}
    s(t) &= \sum_{q=0}^{MN-1}s[q]a\left(t-q\frac{T}{M}\right),
    \label{eq:s_a}
\end{align}
where $s[q]$ is the $q$-th element of $\mathbf{s}$.

\subsection{Time Domain Input-Output Relation}\label{sec:io_time}
Denote the normalized delay and Doppler shifts of the $p$-th path by $l_p$ and $k_p$, respectively, where we have $\tau_p=l_p\frac{T}{M}$ and $\nu_p=k_p\frac{1}{NT}$. Also, define the sets of normalized delay and Doppler shifts by $\mathcal{L}=\{l_1,\dots,l_P\}$ and $\mathcal{K}=\{k_1,\dots,k_P\}$. Here we assume that each path possesses on-grid delay and Doppler shifts, i.e., $l_p,k_p\in\mathbb{Z}$, as we consider an equivalent channel model with on-grid paths
\footnote{
In a physical channel, the ISI and ICI caused by each off-grid path spread over multiple symbols in the DD domain. That is, the sampled channel can be modeled by an equivalent channel with an increased number of resolvable paths, each having on-grid delay and Doppler values \cite{Lin2022OrthogonalModulation,Tong2024ODDM_PhyChan,Lin2023TF2DD,Bello1963CharacterizationChannels}.
}. Then, the discrete channel can be rewritten as
\begin{equation}\label{eq:ddChan_int}
    h[l,k] =
    \begin{cases}
        h_p, & l=l_p, k = k_p \\
        0, &\text{otherwise}
    \end{cases},
\end{equation}
for $l_p\in\mathcal{L},k_p\in\mathcal{K}$. The maximum delay and Doppler shifts are given by $l_{\max}=\max\{\mathcal{L}\}$ and $k_{\max}=\max\{\abs{k_p}|k_p\in\mathcal{K}\}$, respectively.

When a frame-wise cyclic prefix (CP) with a length of $l_{\max}$ is deployed at the transmitter and removed at the receiver, the received signal at each time instance is the superposition of components from $\abs{\mathcal{L}}$ resolvable delay taps. After matched filtering and sampling at $t=q\frac{T}{M}$ for $q=0,\dots,MN-1$, we obtain the time domain received sample vector of ODDM, denoted by $\mathbf{r}\in\mathbb{C}^{MN\times1}$, and its $q$-th element is \cite{Tong2024ODDM_PhyChan}
\begin{equation}\label{eq:io_time}
    r[q] = \sum_{l\in\mathcal{L}} g[l,q]s\bigl[[q-l]_{MN}\bigr] + z[q],
\end{equation}
where $z[q]\sim\mathcal{CN}(0,\sigma_z^2)$ is the sampled AWGN and
\begin{equation}\label{eq:g_q}
    g[l,q] = \sum_{k\in{\mathcal{K}}} h[l,k] e^{j2\pi\frac{k(q-l)}{MN}}
\end{equation}
is the channel impulse response for the $l$-th delay tap at the $q$-th received sample. Here, we define the transmit SNR as $\mathrm{SNR} \triangleq P_t/\sigma_z^2$ with $P_t=\mathbb{E}[\abs{s[q]}^2]$ being the transmit signal power.

\begin{figure}[ht]
    \centering
    \includegraphics[width=8.8cm,trim={0 56 45 62},clip]{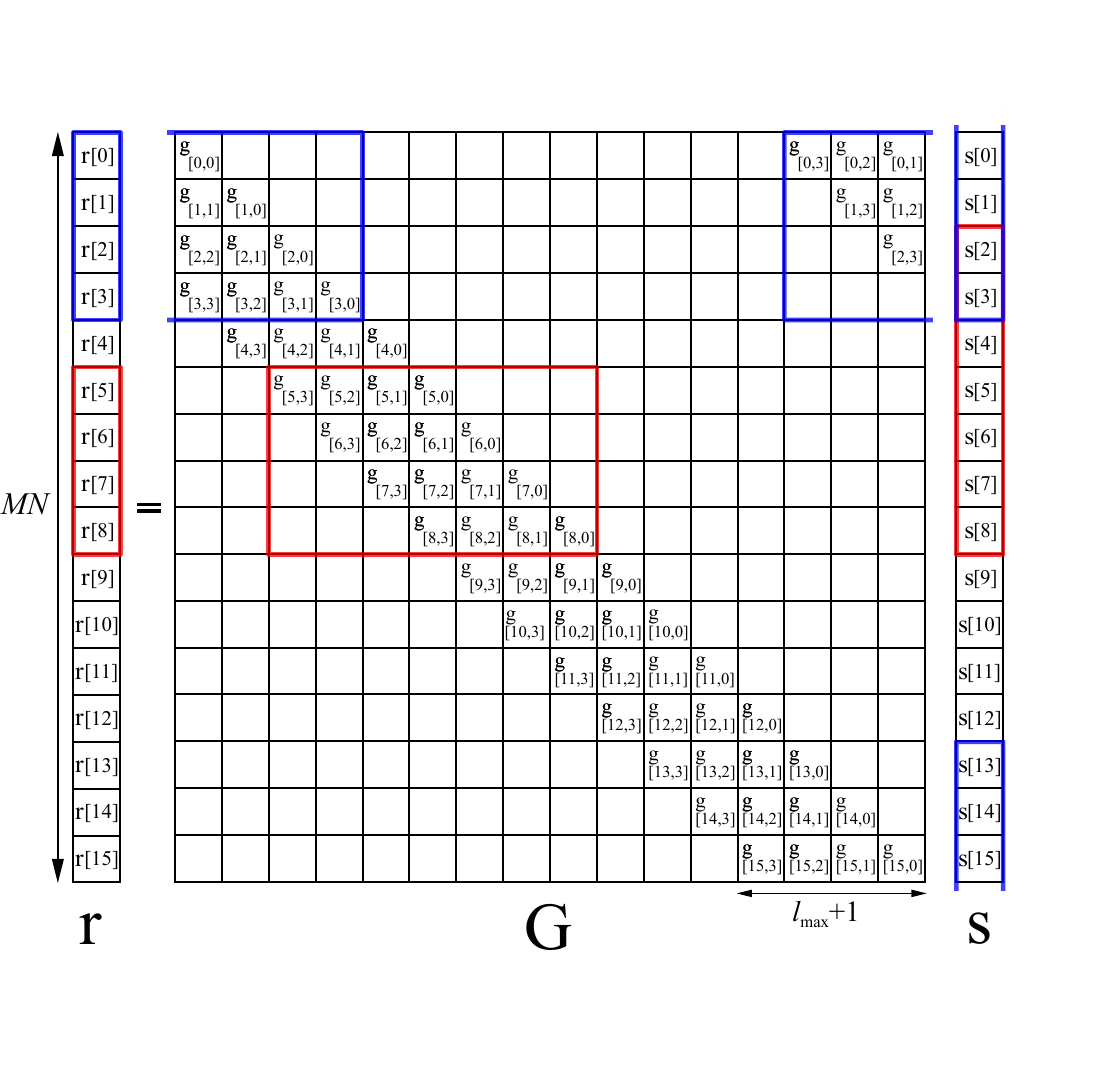}
    \caption{Time domain input-output relation for ODDM, $M=N=4$ and $l_{\max}=3$, with the noise term skipped.}
    \label{fig:io_time}
\end{figure}

Using the time domain transmitted and received sample vectors $\mathbf{s}$ and $\mathbf{r}$, the relation in (\ref{eq:io_time}) can also be written in matrix form as
\begin{equation}\label{eq:io_time_mat}
    \mathbf{r}=\mathbf{G}\mathbf{s}+\mathbf{z},
\end{equation}
where $\mathbf{G}\in\mathbb{C}^{MN\times MN}$ is the time-domain channel matrix whose element at the $q$-th time index and $l$-th delay index is $G\bigl[q,[q-l]_{MN}\bigr] = g[l,q]$ for $q = 0,\dots,MN-1$, $l\in\mathcal{L}$. Fig. \ref{fig:io_time} shows a time domain input-output relation for ODDM with $M=N=4$ and $l_{\max}=3$. We can observe that $\mathbf{G}$ has a quasi-banded structure.

For each transmitted symbol $s[q]$, we can also define a sub-input-output relation to fully capture its \emph{channel impaired} components and interference components. The components of $s[q]$ spreads over received symbols $r\bigl[[q+l]_{MN}\bigr],l\in\mathcal{L}$. Accordingly, we construct the $q$-th sub received symbol vector $\mathbf{r}_q\in\mathbb{C}^{(l_{\max}+1)\times1}$ whose $l$-th element is $r\bigl[[q+l]_{MN}\bigr]$. Meanwhile, $\mathbf{r}_q$ is superimposed by components from $s\bigl[[q+\Delta l]_{MN}\bigr],\Delta l\in\dot{\mathcal{L}}$, where we introduce the time index offset $\Delta l$ of the interfering symbols and its range $\dot{\mathcal{L}}=\{-l_{\max},\dots,0,\dots,l_{\max}\}$. Similarly, we construct the $q$-th sub transmitted symbol vector $\mathbf{s}_q\in\mathbb{C}^{(2l_{\max}+1)\times1}$. Then, we define the truncated spreading vector $\mathbf{g}_{q,\Delta l}\in\mathbb{C}^{(l_{\max}+1)\times1}$ for $s\bigl[[q+\Delta l]_{MN}\bigr]$, where its $l$-th element is given by
\begin{equation}\label{eq:g_q_deltal}
    g_{q,\Delta l}[l] = 
    \begin{cases}
        g\bigl[l-\Delta l,[q + l]_{MN}\bigr], & l-\Delta l\in\mathcal{L},
        \\
        0, & \text{otherwise}.
    \end{cases}
\end{equation}
Hence, the sub-input-output relation for $s[q]$ can be written as
\begin{equation}\label{eq:iosub_time_mat}
    \underbrace{
    \begin{bmatrix}
        r[q]\\
        r\bigl[[q+1]_{MN}\bigr]\\
        \vdots\\
        r\bigl[[q+l_{\max}]_{MN}\bigr]
    \end{bmatrix}
    }_{\mathbf{r}_q}
    =\mathbf{G}_{q}
    \underbrace{
    \begin{bmatrix}
        s\bigl[[q-l_{\max}]_{MN}\bigr]\\
        \vdots\\
        s[q]\\
        \vdots\\
        s\bigl[[q+l_{\max}]_{MN}\bigr]
    \end{bmatrix}
    }_{\mathbf{s}_q}
    +\mathbf{z}_q,
\end{equation}
where $\mathbf{G}_{q} = [\mathbf{g}_{q,-l_{\max}},\dots,\mathbf{g}_{q,0},\dots,\mathbf{g}_{q,l_{\max}}]$ is the sub-channel matrix and $\mathbf{z}_q=\left[z[q],\dots,z\bigl[[q+l_{\max}]_{MN}\bigr]\right]^\tran$ is the corresponding AWGN vector. In Fig. \ref{fig:io_time}, the sub-input-output relations for $s[1]$ and $s[5]$ are illustrated with blue and red boxes, respectively.

Because the sub-input-output relation in (\ref{eq:iosub_time_mat}) uses smaller matrices, it is leveraged by the MRC and SIC-MMSE detectors to enable low-complexity detection. Notably, despite differences in waveform and input-output relation, the time-domain effective channel matrix for OTFS systems also exhibits a quasi-banded structure similar to $\mathbf{G}$ \cite{Tong2024ODDM_PhyChan}. Therefore, MRC and SIC-MMSE also apply to OTFS through corresponding sub-input-output relations. However, in this work, we investigate the performance of these detectors within the ODDM framework, where the ODDM input-output relation is considered.

\subsection{DD Domain Input-Output Relation}\label{sec:iodd}

By employing the inverse operation of vectorization, we can reshape $\mathbf{r}$ and obtain the DT domain received ODDM frame as $\mathbf{Y}^{\mathrm{DT}}=\mathrm{vec}_{M\times N}^{-1}(\mathbf{r})$. Then, $\mathbf{Y}^{\mathrm{DT}}$ can be converted back to DD domain by $N$-point DFT, i.e., $\mathbf{Y} = \mathbf{Y}^{\mathrm{DT}}\mathbf{F}_N$. Regarding the discrete channel in (\ref{eq:ddChan_int}), the DD domain input-output relation of ODDM is derived as \cite{Lin2022OrthogonalModulation,Tong2024ODDM_PhyChan}
\begin{align}
    Y[m,n] & = \sum_{p=1}^P h_p e^{j2\pi\frac{(m-l_p)k_p}{MN}} \alpha_p(m,n)
    \nonumber\\
    &\qquad \times X\bigl[[m-l_p]_M,[n-k_p]_N\bigr] + w[m,n],
    \label{ioddInt}
\end{align}
where 
\begin{equation*}
    \alpha_p(m,n) = 
    \begin{cases}
        1, & m\geq l_p \\
        e^{-j\frac{2\pi[n-k_p]_N}{N}}, & m<l_p
    \end{cases} 
\end{equation*}
characterizes the additional phase rotation for the CP symbols, and $w[m,n]$ is the DD domain noise sample. Because the normalized DFT matrix $\mathbf{F}_N$ is unitary, the noise distribution is conserved in the DD domain. Therefore, $w[m,n]$ adhere to the same distribution as $z[q]$, i.e., $w[m,n]\sim\mathcal{CN}(0,\sigma_z^2)$. The DD domain input-output relation is used for channel estimation.

\section{Channel Estimation}\label{sec:chan_est}

To have tractable performance analysis, we adopt the embedded pilot-aided channel estimation scheme from \cite{Raviteja2019EmbeddedChannels} for ODDM, where full guard symbols are considered. To cover all channel responses, the algorithm assumes a sufficiently large maximum delay spread $l_{\max}$ such that $l_p\leq l_{\max}$ for $p=1,\dots,P$. Based on this scheme, we then establish a model for channel estimation errors. Denote the DD domain pilot symbol in an ODDM frame by $X[m_\mathrm{pilot},n_\mathrm{pilot}]=x_\mathrm{pilot}$, which is placed in the middle of the DD grid at $m_\mathrm{pilot}=M/2, n_\mathrm{pilot}=N/2$. Guard intervals are placed on both the delay and Doppler indices so that the channel-impaired pilot and data symbols do not interfere with each other. The remaining grid points in the frame are used for data transmission. The position of the pilot and guard intervals are known by the receiver.

The channel responses are distributed across a range defined by delay shifts $\hat{\mathcal{L}}=\{0,\dots,l_{\max}\}$ and Doppler shifts $\hat{\mathcal{K}}=\{-N/2,\dots,N/2-1\}$. To contain full channel information, the estimation algorithm uses the symbols $Y[m_\mathrm{pilot}+l,n_\mathrm{pilot}+k]$ for $l\in\hat{\mathcal{L}}$ and $k\in\hat{\mathcal{K}}$. By the input-output relation in (\ref{ioddInt}), the DD channel coefficient for the $p$-th path is estimated as
\begin{equation}\label{eq:chan_est}
    \hat{h}[l_p,k_p]=\frac{Y[m_\mathrm{pilot}+l_p,n_\mathrm{pilot}+k_p]}{x_\mathrm{pilot} e^{j2\pi\frac{m_\mathrm{pilot}k_p}{MN}}}.
\end{equation}
Due to the AWGN, the received frame has non-zero responses for all $Y[m_\mathrm{pilot}+l,n_\mathrm{pilot}+k]$. That is, the estimated channel has non-zero responses at delay shifts $l\in\hat{\mathcal{L}}$ and Doppler shifts $k\in\hat{\mathcal{K}}$. By (\ref{eq:chan_est}), we can model the estimated DD domain channel response as
\begin{equation}\label{eq:hhat}
    \hat{h}[l,k] = h[l,k]+\Delta h[l,k],
\end{equation}
for $l\in\hat{\mathcal{L}},k\in\hat{\mathcal{K}}$, where $\Delta h[l,k] \sim \mathcal{CN}(0,\sigma_z^2/P_\mathrm{pilot}^\mathrm{DD})$ is the channel estimation error and $P_\mathrm{pilot}^\mathrm{DD}=\abs{x_\mathrm{pilot}}^2$ is the DD domain pilot power. By substituting (\ref{eq:hhat}) to (\ref{eq:g_q}), the estimated time domain channel response for the $q$-th time index becomes
\begin{align}\label{eq:ghat}
    \hat{g}[l,q] = g[l,q] + \Delta g[l,q],
\end{align}
for $l\in\hat{\mathcal{L}}$, where
\begin{equation}\label{eq:dg}
    \Delta g[l,q]=\sum_{k\in\hat{\mathcal{K}}}\Delta h[l,k]e^{j2\pi\frac{k(q-l)}{MN}}
\end{equation}
is the time domain channel estimation error. Meanwhile, regarding the definition of $\mathbf{g}_{q,\Delta l}$ in (\ref{eq:g_q_deltal}), we have the estimated truncated spreading vector
\begin{equation}\label{eq:ghat_vec}
    \hat{\mathbf{g}}_{q,\Delta l} = \mathbf{g}_{q,\Delta l} + \Delta\mathbf{g}_{q,\Delta l},
\end{equation}
where the $l$-th element of $\Delta\mathbf{g}_{q,\Delta l}$ is
\begin{equation}\nonumber
    \Delta g_{q,\Delta l}[l] = 
    \begin{cases}
        \Delta g\bigl[l-\Delta l,[q+l]_{MN}\bigr], & l-\Delta l\in\hat{\mathcal{L}},
        \\
        0, & \text{otherwise}.
    \end{cases}
\end{equation}

By (\ref{eq:dg}), it can be seen that the time domain channel estimation error $\Delta g[l,q]$ also follows a complex Gaussian distribution $\mathcal{CN}(0,\sigma_{\Delta g}^2)$, but with variance $\sigma_{\Delta g}^2 = \sigma_z^2 N/P_{\text{pilot}}^\mathrm{DD}$. This indicates that, when the impulse pilot power is defined in the DD domain, the level of the time-domain channel estimation errors depends on the Doppler resolution and scales linearly with $N$. Physically, an impulse pilot in the DD domain spreads to $N$ samples in the time domain, thus experiencing noise from all the $N$ co-delay samples. To get a consistent level of channel estimation error with respect to pilot power, we define the effective pilot power as $P_{\text{pilot}} = P_{\text{pilot}}^\mathrm{DD}/N$, which is essentially the pilot power after time-domain spreading. The effective pilot SNR is thus $\text{SNR}_\text{pilot} = P_{\text{pilot}}/\sigma_z^2$. Considering (\ref{eq:ghat_vec}), the variance of the channel estimation error becomes
\begin{equation}\label{eq:var_dg}
    \var(\Delta g_{q,\Delta l}[l]) = 
    \begin{cases}
        \sigma_{\Delta g}^2, & l-\Delta l\in\hat{\mathcal{L}},
        \\
        0, & \text{otherwise},
    \end{cases}
\end{equation}
where $\sigma_{\Delta g}^2 = \sigma_z^2 /P_{\text{pilot}}$.

With the Gaussian model, the impact of channel estimation error on detection performance is effectively characterized by its variance $\sigma_{\Delta g}^2$. This allows us to analytically evaluate the performance degradation due to imperfect CSI as will be presented in Section \ref{sec:sinr_analysis}.

\section{Detection Algorithms}\label{sec:detectors}

Recall that the time-domain input-output relation of ODDM is characterized by the $MN\times MN$ channel matrix $\mathbf{G}$ in (\ref{eq:io_time_mat}). Instead of direct linear equalization with respect to (\ref{eq:io_time_mat}), MRC and SIC-MMSE leverage the sub-input-output relation in (\ref{eq:iosub_time_mat}) to enable low-complexity detection, where the correlation between consecutive symbol estimates is mitigated by interference cancellation. In what follows, we introduce these two representative detection algorithms.

\subsection{Maximum-Ratio Combining (MRC)}\label{sec:mrc}
The MRC detector directly combines the channel-impaired signal components received at different delay branches. The signal components are obtained by canceling the interference components from the received signal, which requires prior symbol estimates. A TF domain single-tap MMSE equalizer is thus used to initialize the symbol estimates, which helps to facilitate the convergence of the MRC algorithm \cite{Thaj2020LowSystems}.

Denote prior symbol estimates by $\hat{s}[q] = s[q] + \Delta s[q],q=0,\dots,MN-1$, where $\Delta s[q]$ is the symbol estimation error. Define $\tilde{r}_q[l]$ to be the estimated channel-impaired component of $s[q]$ at the $l$-th delay tap. Referring to the input-output relation in (\ref{eq:io_time}), we can find $\tilde{r}_q[l]$ by interference cancellation:
\begin{align}
    \tilde{r}_q[l] &= r\bigl[[q+l]_{MN}\bigr]
    \nonumber\\
    &\qquad-\sum_{\mathclap{l^\prime\neq l,l^\prime\in\mathcal{L}}} g\bigl[l^\prime,[q+l]_{MN}\bigr]\hat{s}\bigl[[q+l-l^\prime]_{MN}\bigr]
    \nonumber\\
    & = g\bigl[l,[q+l]_{MN}\bigr]s[q] + \tilde{z}_q[l],
    \label{eq:rtilde_l}
\end{align}
where
\begin{align}
    \tilde{z}_q[l] &= z\bigl[[q+l]_{MN}\bigr] 
    \nonumber\\
    &\qquad-\sum_{\mathclap{l^\prime\neq l,l^\prime\in\mathcal{L}}} g\bigl[l^\prime,[q+l]_{MN}\bigr]\Delta s\bigl[[q+l-l^\prime]_{MN}\bigr]
    \label{eq:ztilde_l}
\end{align}
is the residual interference plus noise (RIPN) term due to AWGN and imperfect interference cancellation. With (\ref{eq:rtilde_l}) and (\ref{eq:ztilde_l}), we further define the channel impaired branch vector $\tilde{\mathbf{r}}_q=[\tilde{r}_q[0],\dots,\tilde{r}_q[l_{\max}]]^\tran$ and the RIPN vector $\tilde{\mathbf{z}}_q=[\tilde{z}_q[0],\dots,\tilde{z}_q[l_{\max}]]^\tran$. Then, we have
\begin{equation}\label{eq:rtilde_vec}
    \tilde{\mathbf{r}}_q = \mathbf{g}_q s[q] + \tilde{\mathbf{z}}_q,
\end{equation}
where $\mathbf{g}_q =\left[g[0,q],\dots,g\bigl[l_{\max},[q+l_{\max}]_{MN}\bigr]\right]^\tran$ is the time domain channel spreading vector for $s[q]$. By invoking the sub-input-output relation in (\ref{eq:iosub_time_mat}), we can write the RIPN vector as
\begin{equation}\label{eq:ripn}
    \tilde{\mathbf{z}}_q = \mathbf{z}_q -\sum_{\Delta l\neq0,\Delta l\in\dot{\mathcal{L}}} \mathbf{g}_{q,\Delta l}\Delta s\bigl[[q+\Delta l]_{MN}\bigr],
\end{equation}
and it can be shown that $\mathbf{g}_q =\mathbf{g}_{q,0}$.

The MRC detector implicitly assumes perfect interference cancellation \cite{Thaj2020LowSystems}. In such a case, $\tilde{\mathbf{z}}_q$ reduces to AWGN and (\ref{eq:rtilde_vec}) becomes a single input multiple output (SIMO) system without ISI. Then, the MRC output is computed as
\begin{equation}\label{eq:mrc}
    \tilde{s}[q] =\mathbf{w}_q^{\mathrm{MRC}}\tilde{\mathbf{r}}_q,
\end{equation}
where
\begin{align}\label{eq:mrc_coeff}
    \mathbf{w}_q^{\mathrm{MRC}} &= \left(\mathbf{g}_q^\herm\mathbf{g}_q\right)^{-1}\mathbf{g}_q^\herm
\end{align}
is the MRC coefficient. Notably, $\mathbf{w}_q^{\mathrm{MRC}}$ is also a zero-forcing (ZF) coefficient in this case. However, we adopt the name MRC following the original iterative MRC paper \cite{Thaj2020LowSystems} because it better represents the nature of $\mathbf{w}_q^{\mathrm{MRC}}$ as a maximum-ratio combiner assuming zero ISI.

For convenience, we define the time-domain equalized symbol vector at the $m$-th delay as $\tilde{\mathbf{s}}_m\in \mathbb{C}^{N\times1}$, whose $\dot{n}$-th element is $\tilde{s}_m[\dot{n}] = \tilde{s}[\dot{n}M+m]$. Once all the symbols in $\tilde{\mathbf{s}}_m$ have been equalized, we can get the DD domain soft-decision estimates $\tilde{\mathbf{x}}_m$ by $N$-point DFT:
\begin{equation}\label{eq:dft_sm2xm}
    \tilde{\mathbf{x}}_m = \mathbf{F}_N\tilde{\mathbf{s}}_m.
\end{equation}
The corresponding hard-decision estimates are obtained via element-wise ML detection:
\begin{equation}\label{eq:ml}
    \hat{x}_m[n] = \argmin_{a_j\in\Lambda}{|a_j-\tilde{x}_m[n]|},
\end{equation}
where $\Lambda=\{a_1,\dots,a_A\}$ denotes an $A$-ary constellation set.
Then, $\hat{\mathbf{x}}_m$ can be converted back to time domain by IDFT:
\begin{equation}\label{eq:idft_xm2sm}
    \hat{\mathbf{s}}_m = \mathbf{F}_N^\herm\hat{\mathbf{x}}_m,
\end{equation}
where the $\dot{n}$-th element of $\hat{\mathbf{s}}_m\in\mathbb{C}^{N\times1}$ is $\hat{s}[\dot{n}M+m]$. This completes the symbol estimation at delay index $m$.

\begin{remark}
    In operations (\ref{eq:dft_sm2xm})-(\ref{eq:idft_xm2sm}), the gain in error performance is achieved through the element-wise ML detection in (\ref{eq:ml}) \cite[Prop. 2]{Li2022CrossModulation}. However, the domain conversion by DFT and IDFT are collectively performed on a group of $N$ symbols. To progressively leverage the ML gain, in this paper, we explore a cross-domain SIC scheme for both MRC and SIC-MMSE detectors, as defined in Definition \ref{definition:sic}.
\end{remark}

\begin{definition}[Cross-domain SIC scheduling]\label{definition:sic}
    Define $m_0\in[0,M-1]$ to be the delay index where the SIC process starts. Interference cancellation is performed sequentially for each delay index, proceeding one at a time for $m=[m_0+\Delta m]_{M},\Delta m=0,\dots,M-1$. At each delay index $m$, parallel interference cancellation and equalization are performed to obtain the post-equalization symbols $\tilde{\mathbf{s}}_m$. Then, cross-domain ML detection is performed on $\tilde{\mathbf{s}}_m$ to get the updated symbol estimates $\hat{\mathbf{s}}_m$ which are immediately used in the interference cancellation for $\tilde{\mathbf{s}}_{[m+l]_M},l\in\mathcal{L}$.
\end{definition}

To improve error performance, steps (\ref{eq:rtilde_l})-(\ref{eq:idft_xm2sm}) are run iteratively. The detection results from the previous iteration will be used as the initial estimates of symbols in the next iteration. Since the symbol estimates are iteratively updated, the channel impaired branch vector $\tilde{\mathbf{r}}_q$ needs to be re-estimated accordingly. Instead of directly using (\ref{eq:rtilde_l}), we apply the low-complexity method described in \cite{Thaj2020LowSystems} to compute $\tilde{\mathbf{r}}_q$ within $2\abs{\mathcal{L}}$ complex multiplications. Considering (\ref{eq:mrc}) and the pair of FFT and IFFT for every $N$ symbols, the complexity order of the iterative MRC detector is $\mathcal{O}(n_\mathrm{ite}MN(\log_2N+\abs{\mathcal{L}}))$, where $n_\mathrm{ite}$ is the number of iterations.

\subsection{Hard SIC-MMSE}\label{sec:hardsicmmse}
The SIC-MMSE detector uses the sub-input-output relation in (\ref{eq:iosub_time_mat}) to reduce the dimension of MMSE filtering. Similar to MRC, prior symbol estimates are used to perform interference cancellation and get $\tilde{\mathbf{r}}_q$ in (\ref{eq:rtilde_vec}), where the cross-domain SIC scheduling in Definition \ref{definition:sic} is followed. The MMSE output for the $q$-th time domain symbol is
\begin{equation}\label{eq:sicmmse}
    \tilde{s}[q] = \mathbf{w}_q^{\mathrm{MMSE}}\tilde{\mathbf{r}}_q/\mu_q,
\end{equation}
where
\begin{equation}\label{eq:mmse_filter}
    \mathbf{w}_q^{\mathrm{MMSE}} = \mathbf{g}_q^\herm\left(\mathbf{G}_q\mathbf{V}_q\mathbf{G}_q^\herm+\sigma_z^2\right)^{-1},
\end{equation}
is the MMSE filter, and $\mu_q=\mathbf{w}_q^{\mathrm{MMSE}}\mathbf{g}_q$ is the normalization factor. We have $\mathbf{V}_q=\mathbb{E}\left[\Delta\mathbf{s}_q\Delta\mathbf{s}_q^\herm\right]$ being the covariance matrix of the symbol estimates. Since the time-domain transmitted symbols $\mathbf{s}$ are i.i.d., $\mathbf{V}_q$ is a diagonal matrix.

In the hard SIC-MMSE, the hard-decision estimate by (\ref{eq:ml}) is employed as the symbol estimate $\hat{\mathbf{x}}_m$. Due to the lack of prior information, all symbols are initialized as zeros with an error variance of $P_t$. As the detection is performed following the SIC scheduling in Definition \ref{definition:sic}, previously estimated symbols are treated as perfectly canceled with an error variance of zero. In this way, the hard-decision covariance matrix, denoted by $\mathbf{V}_q^{\text{hard}}$, evolves with respect to the delay index $m$.

The SIC-MMSE algorithm is run iteratively to improve detection accuracy. For iteration $i>1$, because we have already obtained prior estimates of all transmitted symbols and assume perfect interference cancellation, we have a constant covariance matrix. In particular,
\begin{equation}\label{eq:cov_hard_i>1}
    \mathbf{V}_q^{\text{hard},(i)}=\diag([0,\dots,0,P_t,0,\dots,0]),
\end{equation}
for all time indices $q$ with $i>1$. By substituting (\ref{eq:cov_hard_i>1}) to (\ref{eq:mmse_filter}), the MMSE filter for iteration $i>1$ can be simplified into a form akin to that of the MRC coefficient, given by
\begin{align}
    \mathbf{w}_q^{\text{hard},(i)} =\left(P_t\mathbf{g}_q^\herm\mathbf{g}_q+\sigma_z^2\right)^{-1}\mathbf{g}_q^\herm,
    \label{eq:hardmmse_filter_>1}
\end{align}
which avoids matrix inversion.

Same as MRC, SIC-MMSE updates $\tilde{\mathbf{r}}_q$ using the low-complexity method described in \cite{Thaj2020LowSystems} for computational efficiency. Since matrix inversion is only involved in the first iteration, the overall complexity of hard SIC-MMSE is $\mathcal{O}\left(n_\mathrm{ite}MN(\log_2N+\abs{\mathcal{L}})+MN(l_{\max}+1)^3\right)$. Empirically, the algorithm converges within five iterations as will be shown in Section \ref{sec:numerical_results}.

\subsection{Soft SIC-MMSE}\label{sec:softsicmmse}
Instead of assuming perfect interference cancellation, soft SIC-MMSE takes into account the propagation of decision errors. Following the cross-domain SIC scheduling in Definition \ref{definition:sic}, the a posteriori error variances of previous symbol estimates are used as the a priori error variances in subsequent estimations to enable better detection accuracy. Thus, the soft SIC-MMSE filter is
\begin{equation}\label{eq:softmmse_filter}
    \mathbf{w}_q^{\mathrm{soft}} = \mathbf{g}_q^\herm\left(\mathbf{G}_q\mathbf{V}_q^{\text{soft}}\mathbf{G}_q^\herm+\sigma_z^2\right)^{-1},
\end{equation}
where
\begin{align}
    \mathbf{V}_q^{\text{soft}}=\diag\Bigl(\Bigl[\var\bigl(\Delta s\bigl[[q-&l_{\max}]_{MN}\bigr]\bigr),\dots,
    \nonumber\\
    &\var\bigl(\Delta s\bigl[[q+l_{\max}]_{MN}\bigr]\bigr)\Bigr]\Bigr)
    \nonumber
\end{align}
with $\var(\Delta s[q])=P_t$.

As in the hard decision scheme, the error variances of all the symbol estimates are initialized to be $P_t$. After SIC-MMSE detection, we want to approximate the error variances of the updated symbol estimates so that they can be used in subsequent symbol estimation. Upon substituting (\ref{eq:rtilde_vec}) and (\ref{eq:softmmse_filter}) to (\ref{eq:sicmmse}), the MMSE output is written as
\begin{equation}\label{eq:stilde_soft}
    \tilde{s}[q] = \left(\mathbf{w}_q^{\mathrm{soft}}\mathbf{g}_q s[q] + \mathbf{w}_q^{\mathrm{soft}}\tilde{\mathbf{z}}_q\right)/\mu_q^{\mathrm{soft}},
\end{equation}
where $\mu_q^{\mathrm{soft}}=\mathbf{w}_q^{\mathrm{soft}}\mathbf{g}_q$ is the corresponding normalization factor. When the number of resolvable paths is sufficiently large, the MMSE-suppressed noise term $\mathbf{w}_q^{\mathrm{soft}}\tilde{\mathbf{z}}_q$ is approximately Gaussian \cite{Poor1997ProbabilityDetection}. Therefore, the time-domain post-MMSE variance can be approximated as $\var(\Delta\tilde{s}[q]) = P_t\left(1-\mu_q^{\mathrm{soft}}\right)/\left(\mu_q^{\mathrm{soft}}\right)$ \cite{Li2024SICMMSE_Turbo}.

With the MMSE output, soft-decision estimates of the symbols are computed by incorporating the constellation constraints. As in MRC and hard SIC-MMSE, we firstly obtain the DD domain observation $\tilde{\mathbf{x}}_m$ by DFT. Because the transformation by DFT is unitary, the average post-MMSE variance is preserved in the DD domain. Moreover, when $N$ becomes large enough, the variances of $\Delta\tilde{x}_m[n],n=0,\dots,N-1$ approaches the same value due to the law of large numbers \cite{Li2022CrossModulation}. Based on this assumption, we have
\begin{equation}\label{eq:vardx=vards}
    \var(\Delta\tilde{x}_m[n]) = \mathbb{E}[\var(\Delta\tilde{s}_m[\dot{n}])],
\end{equation}
for $n=0,\dots,N-1$.

Therefore, we can perform symbol-by-symbol soft-decision symbol estimation in the DD domain. Regarding the constellation constraints, we have
\begin{equation}
    \Pr\left(\tilde{x}_m[n]\middle|x_m[n]=a_j\right) = \frac{\exp\left(\frac{\abs{\tilde{x}_m[n]-a_j}^2}{\var(\Delta\tilde{x}_m[n])}\right)}{\sum_{a_k\in\Lambda} \exp\left(\frac{\abs{\tilde{x}_m[n]-a_k}^2}{\var(\Delta\tilde{x}_m[n])}\right)}.
\end{equation}
Then, the a posteriori estimate of $x_m[n]$ is computed as
\begin{align}
    \hat{x}_m^\text{soft}[n] = \sum_{a_j\in\Lambda} \Pr\left(\tilde{x}_m[n]\middle|x_m[n]=a_j\right)\times a_j,
\end{align}
with the corresponding a posteriori variance
\begin{align}
    \var(&\Delta x_m[n])
    \nonumber\\
    &=\sum_{a_j\in\Lambda} \Pr\left(\tilde{x}_m[n]\middle|x_m[n]=a_j\right)\times \abs{a_j-\hat{x}_m^\text{soft}[n]}^2.
\end{align}
The variance forms the diagonal entries of the corresponding DD domain a posteriori covariance matrix $\mathbf{V}_m^{\mathrm{DD}}$. Meanwhile, by the i.i.d. assumption on $\mathbf{x}_m$, the non-diagonal entries in $\mathbf{V}_m^{\mathrm{DD}}$ are simply zero.

After getting the DD domain a posteriori estimates $\hat{\mathbf{x}}_m^\text{soft}$, it can be directly converted back to time domain symbols $\hat{\mathbf{s}}_m$ by IDFT. The corresponding time domain covariance matrix can be obtained by $\mathbf{V}_m^{\mathrm{DT}} = \mathbf{F}_N^\herm\mathbf{V}_m^{\mathrm{DD}}\mathbf{F}_N$. In subsequent symbol estimation, the a posteriori estimates $\hat{\mathbf{s}}_m$ and the diagonal entries of $\mathbf{V}_m^{\mathrm{DT}}$ will be used as a priori estimates and error variances to perform SIC-MMSE equalization \cite{Guo2011ADetector}.

Soft SIC-MMSE also adopts the cross-domain SIC scheduling in Definition \ref{definition:sic} and runs iteratively to improve detection accuracy. Because the covariance matrix $\mathbf{V}_q$ is updated after every iteration, the soft-decision MMSE filter $\mathbf{w}_q^{\text{soft}}$ needs to be recomputed by (\ref{eq:softmmse_filter}) in every iteration. And the associated matrix inversion becomes a dominant source of computational complexity. Therefore, the overall complexity of the soft SIC-MMSE algorithm is $\mathcal{O}\left(n_\mathrm{ite}MN\left(\log_2N+(l_{\max}+1)^3\right)\right)$.

\section{Signal-to-Interference-Plus-Noise Ratio Analysis}\label{sec:sinr_analysis}

In this section, we analyze the post-equalization SINR of iterative MRC and SIC-MMSE, incorporating the cross-domain SIC scheduling as defined in Definition \ref{definition:sic}. We derive their $i$-th iteration post-equalization SINR in the presence of channel estimation errors, where we prove that hard SIC-MMSE is equivalent to MRC from the second iteration onwards. An SINR upper bound is provided to demonstrate the robustness of these linear detectors against channel estimation errors. Our analytical results are verified through simulations, which extend the current understanding of interference cancellation-based iterative detectors under imperfect CSI and inspire ODDM detector designs. This SINR analysis provides a foundation for the subsequent BER analysis in Section \ref{sec:ber&improve}.

\subsection{Post-Equalization SINR in the $i$-th Iteration}\label{sec:post_eq_sinr_i}
For both MRC and SIC-MMSE, denote the symbol estimate in the $i$-th iteration as
\begin{equation}
    \hat{s}^{(i)}[q] = s[q] + \Delta s^{(i)}[q],
\end{equation}
for $q=0,\dots,MN-1$, where $\Delta s^{(i)}[q]$ is the corresponding symbol estimation error. The distribution of $\Delta s^{(i)}[q]$ is typically non-Gaussian in hard decision schemes due to the nonlinear ML detection across iterations. Here we only assume $\Delta s^{(i)}[q]$ has zero mean and a known variance $\left(\sigma_e^2\right)^{(i)}$.

To model the signal before equalization, we first derive the interference cancellation step for both MRC and SIC-MMSE. Following the cross-domain SIC scheduling in Definition \ref{definition:sic} and the evolution of symbol estimates across iterations, there are three cases for the available symbol estimates in computing $\tilde{\mathbf{s}}_m$ with $m=[m_0+\Delta m]_M$:
\begin{itemize}
    \item Case 1: When $\Delta m<l_{\max}$, last-iteration estimates $\hat{\mathbf{s}}_{[m+\Delta l]_{M}}^{(i-1)}$ are available if $\Delta l>0$ or $\Delta l<-\Delta m$ while current-iteration estimates $\hat{\mathbf{s}}_{[m+\Delta l]_{M}}^{(i)}$ are available if $-\Delta m\leq\Delta l<0$.
    \item Case 2: When $l_{\max}\leq\Delta m< M-l_{\max}$, last-iteration estimates $\hat{\mathbf{s}}_{[m+\Delta l]_{M}}^{(i-1)}$ are available if $\Delta l>0$ while current-iteration estimates $\hat{\mathbf{s}}_{[m+\Delta l]_{M}}^{(i)}$ are available if $\Delta l<0$.
    \item Case 3: When $\Delta m\leq M-l_{\max}$, last-iteration estimates $\hat{\mathbf{s}}_{[m+\Delta l]_{M}}^{(i-1)}$ are available if $0<\Delta l<M-\Delta m$ while current-iteration estimates $\hat{\mathbf{s}}_{[m+\Delta l]_{M}}^{(i)}$ are available if $\Delta l<0$ or $\Delta l\geq M-\Delta m$.
\end{itemize}
It can be seen that when we have $M\gg l_{\max}$, the majority of delay indices fall in Case 2. For simplicity, in this analysis, we consider a sufficiently large $M$ and assume the available symbol estimates comply with Case 2 for the interference cancellation of all symbols.

By considering Case 2 and substituting (\ref{eq:ghat_vec}) to (\ref{eq:rtilde_vec}), the $i$-th iteration interference cancellation result with errors is
\begin{align}
    \tilde{\mathbf{r}}_q^{(i)} = \mathbf{g}_q s[q]+\hat{\mathbf{z}}_q^{(i)},
    \label{eq:rtilde_err}
\end{align}
where
\begin{align}
    \hat{\mathbf{z}}_q^{(i)} = \mathbf{z}_q &-\sum_{\mathclap{\Delta l<0,\Delta l\in\dot{\mathcal{L}}}} \hat{\mathbf{g}}_{q,\Delta l}\Delta s^{(i)}\bigl[[q+\Delta l]_{MN}\bigr]
    \nonumber\\
    &-\sum_{\mathclap{\Delta l>0,\Delta l\in\dot{\mathcal{L}}}} \hat{\mathbf{g}}_{q,\Delta l}\Delta s^{(i-1)}\bigl[[q+\Delta l]_{MN}\bigr]
    \nonumber\\
    &-\sum_{\mathclap{\Delta l\neq0,\Delta l\in\dot{\mathcal{L}}}} \Delta\mathbf{g}_{q,\Delta l} s\bigl[[q+\Delta l]_{MN}\bigr]
    \label{eq:z_hat}
\end{align}
is the RIPN vector with channel estimation errors.

\subsubsection{MRC}
Having derived the symbols after interference cancellation, we now analyze the post-equalization SINR. Consider the MRC coefficient with imperfect CSI
\begin{align}
    \hat{\mathbf{w}}_q^{\mathrm{MRC}} &= \bigl(\underbrace{\hat{\mathbf{g}}_q^\herm\hat{\mathbf{g}}_q}_{\hat{v}_q^\mathrm{MRC}}\bigr)^{-1}\hat{\mathbf{g}}_q^\herm,
    \label{eq:mrc_filter_err}
\end{align}
where $\hat{v}_q^\mathrm{MRC}$ is the normalization factor with channel estimation errors. By substituting (\ref{eq:rtilde_err}) and (\ref{eq:mrc_filter_err}) to the MRC output in (\ref{eq:mrc}), we have
\begin{align}
    \tilde{s}^{(i)}[q] &=\hat{\mathbf{w}}_q^{\mathrm{MRC}}\tilde{\mathbf{r}}_q^{(i)}
    \nonumber\\
    &= \Bigl(\underbrace{\hat{\mathbf{g}}_q^\herm\hat{\mathbf{g}}_qs[q]}_{\psi_q^{\mathrm{MRC}}}+\underbrace{\hat{\mathbf{g}}_q^\herm\hat{\mathbf{z}}^{(i)}_q-\hat{\mathbf{g}}_q^\herm\Delta\mathbf{g}_qs[q]}_{\eta_q^{\mathrm{MRC},(i)}}\Bigr)/\hat{v}_q^\mathrm{MRC},
    \label{eq:s_tilde_mrc}
\end{align}
where $\psi_q^{\mathrm{MRC}}$ represents signal component while $\eta_q^{\mathrm{MRC},(i)}$ represents noise plus interference component. Therefore, we have the following proposition:

\begin{proposition}\label{proposition:sinr_mrc}
    Given a channel realization and the variances of channel and symbol estimation errors, for the $q$-th symbol, the post-equalization SINR by MRC in the $i$-th iteration is
    \begin{equation}\label{eq:sinr_analytical}
        \mathrm{SINR}_q^{\mathrm{MRC},(i)} = \frac{\mathbb{E}\left[\psi^{\mathrm{MRC}}[q]\left(\psi^{\mathrm{MRC}}[q]\right)^*\right]}{\mathbb{E}\left[\eta^{\mathrm{MRC},(i)}[q]\left(\eta^{\mathrm{MRC},(i)}[q]\right)^*\right]},
    \end{equation}
    where
    \begin{align}
        \mathbb{E}\Bigl[\psi_q^{\mathrm{MRC}}&\left(\psi_q^{\mathrm{MRC}}\right)^*\Bigr] = P_t\mathbb{E}\Bigl[\left(\hat{\mathbf{g}}_q^\herm\hat{\mathbf{g}}_q\right)^2\Bigr]
        \nonumber\\
        &= P_t\Bigl(\bigl(\mathbf{g}_q^\herm\mathbf{g}_q\bigr)^2+2\mathbf{g}_q^\herm\mathbf{g}_q\mathbb{E}\bigl[\Delta\mathbf{g}_q^\herm\Delta\mathbf{g}_q\bigr]
        \nonumber\\
        &\quad+\mathbb{E}\bigl[\mathbf{g}_q^\herm\Delta\mathbf{g}_q\Delta\mathbf{g}_q^\herm\mathbf{g}_q\bigr] + \mathbb{E}\bigl[\Delta\mathbf{g}_q^\herm\mathbf{g}_q\mathbf{g}_q^\herm\Delta\mathbf{g}_q\bigr]
        \nonumber\\
        &\quad+\mathbb{E}\Bigl[\bigl(\Delta\mathbf{g}_q^\herm\Delta\mathbf{g}_q\bigr)^2\Bigr]\Bigr),
        \label{eq:psi_mrc}
    \end{align}
    and
    \begin{align}
        \mathbb{E}\left[\eta_q^{\mathrm{MRC},(i)}\left(\eta_q^{\mathrm{MRC},(i)}\right)^*\right] &= \mathbb{E}\biggl[\mathbf{g}_q^\herm\hat{\mathbf{z}}^{(i)}_q\Bigl(\hat{\mathbf{z}}^{(i)}_q\Bigr)^\herm\mathbf{g}_q\biggr]
        \nonumber\\
        &\quad+\mathbb{E}\biggl[\Delta\mathbf{g}_q^\herm\hat{\mathbf{z}}^{(i)}_q\Bigl(\hat{\mathbf{z}}^{(i)}_q\Bigr)^\herm\Delta\mathbf{g}_q\biggr]
        \nonumber\\
        &\quad+P_t\mathbb{E}\bigl[\mathbf{g}_q^\herm\Delta\mathbf{g}_q\Delta\mathbf{g}_q^\herm\mathbf{g}_q\bigr]
        \nonumber\\
        &\quad+P_t\mathbb{E}\Bigl[\bigl(\Delta\mathbf{g}_q^\herm\Delta\mathbf{g}_q\bigr)^2\Bigr],
        \label{eq:eta_mrc}
    \end{align}
    with $\hat{\mathbf{z}}^{(i)}_q$ defined in (\ref{eq:z_hat}).
\end{proposition}
\begin{IEEEproof}
    The detailed derivations of the terms in (\ref{eq:psi_mrc}) and (\ref{eq:eta_mrc}) are given in Appendix \ref{appendix:psi_eta}.
\end{IEEEproof}

\subsubsection{Hard SIC-MMSE}
As discussed in Section \ref{sec:hardsicmmse}, the equalization filter of hard SIC-MMSE has a varying pattern in the first iteration. For simplicity, here we focus on the SINR performance of hard SIC-MMSE in the second and subsequent iterations, with the corresponding filter given in (\ref{eq:hardmmse_filter_>1}). Upon substituting $\hat{\mathbf{g}}_q$ for $\mathbf{g}_q$, the hard SIC-MMSE filter for $i>1$ with imperfect CSI becomes
\begin{align}
    \hat{\mathbf{w}}_q^{\text{hard},(i)} = \bigl(\underbrace{P_t\hat{\mathbf{g}}_q^\herm\hat{\mathbf{g}}_q+\sigma_z^2}_{\hat{v}_q^{\text{hard}}}\bigr)^{-1}\hat{\mathbf{g}}_q^\herm,
    \label{eq:hardmmse_filter_>1_hat}
\end{align}
where the matrix inversion is also encapsulated by the reciprocal of a scalar $\hat{v}_q^{\text{hard}}$. By substituting (\ref{eq:rtilde_err}) and (\ref{eq:hardmmse_filter_>1_hat}) into the SIC-MMSE output in (\ref{eq:sicmmse}), we have
\begin{align}
    \tilde{s}^{(i)}[q] &=\hat{\mathbf{w}}_q^{\text{hard},(i)}\tilde{\mathbf{r}}_q^{(i)}/\hat{\mu}_q^{\text{hard}}
    \nonumber\\
    &=\Bigl(\underbrace{\hat{\mathbf{g}}_q^\herm\hat{\mathbf{g}}_qs[q]}_{\psi_q^{\mathrm{hard}}}+\underbrace{\hat{\mathbf{g}}_q^\herm\hat{\mathbf{z}}^{(i)}_q-\hat{\mathbf{g}}_q^\herm\Delta\mathbf{g}_qs[q]}_{\eta_q^{\mathrm{hard},(i)}}\Bigr)/\left(\hat{\mu}_q^{\text{hard}}\hat{v}_q^{\mathrm{hard}}\right),
    \label{eq:s_tilde_hardsicmmse}
\end{align}
where $\hat{\mu}_q^{\text{hard}} = \hat{\mathbf{w}}_q^{\text{hard},(i)}\hat{\mathbf{g}}_q$ represents the normalization factor with channel estimation errors, $\psi_q^{\mathrm{hard}}$ represents signal component, and $\eta_q^{\mathrm{hard},(i)}$ represents noise plus interference component. Comparing (\ref{eq:s_tilde_hardsicmmse}) with the MRC output in (\ref{eq:s_tilde_mrc}), we can observe that, the hard SIC-MMSE output only differs by a scaling factor, whereby we establish the following proposition:

\begin{proposition}\label{proposition:sinr_hard=sinr_mrc}
    Given the same prior symbol estimates and channel estimation errors, the post-equalization SINR of hard SIC-MMSE from the second and following iterations equals that of MRC, which is
    \begin{equation}\label{eq:sinr_hard=mrc}
        \mathrm{SINR}_q^{\mathrm{hard},(i)} = \mathrm{SINR}_q^{\mathrm{MRC},(i)},
    \end{equation}
    for $i>1$.
\end{proposition}
\begin{IEEEproof}
    The proof follows similar to (\ref{eq:sinr_analytical}) to (\ref{eq:eta_mrc}).
\end{IEEEproof}

\begin{remark}\label{remark:1st_ite}
    It should be noted that Proposition \ref{proposition:sinr_hard=sinr_mrc} does not imply identical error performance between MRC and hard SIC-MMSE. As will be shown in Section \ref{sec:convergence}, the first iteration of hard SIC-MMSE can provide better initial symbol estimates compared to MRC, which contributes to lower BER upon convergence.
\end{remark}

As stated in Section \ref{sec:mrc}, the sub-input-output relation in (\ref{eq:rtilde_vec}) is regarded as a SIMO system when considering hard-decision interference cancellation. In light of this, Proposition \ref{proposition:sinr_hard=sinr_mrc} can be intuitively understood as that there is no ISI in a SIMO system, where MMSE is equivalent to MRC.

\subsubsection{Soft SIC-MMSE}
By substituting $\hat{\mathbf{G}}_q$ to (\ref{eq:softmmse_filter}), the soft SIC-MMSE filter in the $i$-th iteration with channel estimation errors is
\begin{align}
    \hat{\mathbf{w}}_q^{\mathrm{soft},(i)} &= \hat{\mathbf{g}}_q^\herm\left(\hat{\mathbf{G}}_q\mathbf{V}_q^{\mathrm{soft},(i)}\hat{\mathbf{G}}_q^\herm+\sigma_z^2\mathbf{I}\right)^{-1}.
    \label{eq:what_soft}
\end{align}
However, the matrix inverse cannot be encapsulated by a scalar as in (\ref{eq:hardmmse_filter_>1_hat}). Let $\mathbf{K}_q = \mathbf{G}_q\mathbf{V}_q\mathbf{G}_q^\herm+\sigma_z^2\mathbf{I}$ and drop the iteration index. (\ref{eq:what_soft}) can be rewritten as
\begin{align}
    \hat{\mathbf{w}}_q = (\mathbf{g}_q+\Delta\mathbf{g}_q)^\herm(\mathbf{K}_q+\Delta\mathbf{K}_q)^{-1},
\end{align}
where $\Delta\mathbf{K}_q=\mathbf{G}_q\mathbf{V}_q\Delta\mathbf{G}_q^\herm+\Delta\mathbf{G}_q\mathbf{V}_q\mathbf{G}_q^\herm+\Delta\mathbf{G}_q\mathbf{V}_q\Delta\mathbf{G}_q^\herm$ can be viewed as a small perturbation to $\mathbf{K}_q$. If we can approximate $(\mathbf{K}_q+\Delta\mathbf{K}_q)^{-1}$ by a linear function of $\Delta\mathbf{g}_q$ and $\Delta\mathbf{G}_q$, the post-equalization SINR of soft SIC-MMSE under imperfect CSI can be derived in a linear regime. For general MMSE estimators, a first-order approximation by the Neumann series can be made as \cite{Eraslan2013PerformanceError}
\begin{equation}\label{eq:K_inv_approx}
    (\mathbf{K}_q+\Delta\mathbf{K}_q)^{-1}\approx \mathbf{K}_q^{-1}-\mathbf{K}_q^{-1}\Delta\mathbf{K}_q\mathbf{K}_q^{-1}.
\end{equation}
However, the approximation in (\ref{eq:K_inv_approx}) is not tight for SIC-MMSE because $\mathbf{K}_q$ can be ill-conditioned.

Therefore, we perform the SINR analysis for soft SIC-MMSE in the absence of channel estimation error. By following (\ref{eq:stilde_soft}) and incorporating the iteration index, we have the soft-decision SIC-MMSE output
\begin{equation}\label{eq:s_tilde_softsicmmse}
    \tilde{s}^{(i)}[q] = \Bigl(\underbrace{\mathbf{w}_q^{\mathrm{soft},(i)}\mathbf{g}_qs[q]}_{\psi^{\mathrm{soft},(i)}[q]}+\underbrace{\mathbf{w}_q^{\mathrm{soft},(i)}\tilde{\mathbf{z}}_q^{(i)}}_{\eta^{\mathrm{soft},(i)}[q]}\Bigr)/\mu_q^{\mathrm{soft},(i)},
\end{equation}
where $\mu_q^{\mathrm{soft},(i)}=\mathbf{w}_q^{\mathrm{soft},(i)}\mathbf{g}_q$ is the corresponding normalization factor, $\psi^{\mathrm{soft},(i)}[q]$ represents signal component, and $\eta^{\mathrm{soft},(i)}[q]$ represents noise plus interference component. And the $i$-th iteration RIPN vector $\tilde{\mathbf{z}}_q$ without channel estimation error is given as
\begin{align}
    \tilde{\mathbf{z}}_q^{(i)} = \mathbf{z}_q &-\sum_{\mathclap{\Delta l<0,\Delta l\in\dot{\mathcal{L}}}} \mathbf{g}_{q,\Delta l}\Delta s^{(i)}\bigl[[q+\Delta l]_{MN}\bigr]
    \nonumber\\
    &-\sum_{\mathclap{\Delta l>0,\Delta l\in\dot{\mathcal{L}}}} \mathbf{g}_{q,\Delta l}\Delta s^{(i-1)}\bigl[[q+\Delta l]_{MN}\bigr].
\end{align}
Hence, we have the following proposition:
\begin{proposition}\label{proposition:sinr_softsicmmse}
    Given a channel realization and the variances of symbol estimation errors, for the $q$-th symbol, the post-equalization SINR by soft SIC-MMSE in the $i$-th iteration is
    \begin{equation}\label{eq:sinr_soft_analytical}
        \mathrm{SINR}_q^{\mathrm{soft},(i)} = \frac{\mathbb{E}\left[\psi^{\mathrm{soft},(i)}[q]\left(\psi^{\mathrm{soft},(i)}[q]\right)^*\right]}{\mathbb{E}\left[\eta^{\mathrm{soft},(i)}[q]\left(\eta^{\mathrm{soft},(i)}[q]\right)^*\right]},
    \end{equation}
    where
    \begin{equation}
        \mathbb{E}\left[\psi^{\mathrm{soft},(i)}[q]\left(\psi^{\mathrm{soft},(i)}[q]\right)^*\right] = P_t\abs{\mathbf{w}_q^{\mathrm{soft},(i)}\mathbf{g}_q}^2,
    \end{equation}
    and
    \begin{align}
        \mathbb{E}\Bigl[\eta^{\mathrm{soft},(i)}[q]\Bigl(&\eta^{\mathrm{soft},(i)}[q]\Bigr)^*\Bigr]= \sigma_z^2\norm{\mathbf{w}_q^{\mathrm{soft},(i)}}^2
        \nonumber\\
        &\quad+(\sigma_e^2)^{(i)}\sum_{\mathclap{\Delta l<0,\Delta l\in\dot{\mathcal{L}}}}\abs{\mathbf{w}_q^{\mathrm{soft},(i)}\mathbf{g}_{q,\Delta l}}^2
        \nonumber\\
        &\quad+(\sigma_e^2)^{(i-1)}\sum_{\mathclap{\Delta l>0,\Delta l\in\dot{\mathcal{L}}}}\abs{\mathbf{w}_q^{\mathrm{soft},(i)}\mathbf{g}_{q,\Delta l}}^2.
    \end{align}
\end{proposition}
\begin{IEEEproof}
    The proof is straightforward by using (\ref{eq:s_tilde_softsicmmse}) and following similar steps as in Proposition \ref{proposition:sinr_mrc}.
\end{IEEEproof}

\subsection{The impacts of error variance on the theoretical SINR}\label{sec:sinr_errvar_impact}
To verify the accuracy of the derived SINR in Section \ref{sec:post_eq_sinr_i}, we first examine our analytical equations for the $i$-th iteration SINR by using the practical error variance extracted from the simulation data. We use the same ODDM parameters for numerical results as summarized in Table \ref{tab:param}. We first find the per-iteration mean squared error (MSE) of the simulated symbol estimates. For a time-domain ODDM sequence $s[q],q=0,\dots,MN-1$, the MSE of symbol estimates in the $i$-th iteration is computed as
\begin{equation}\label{eq:mse}
    \mse_{s}^{(i)} = \mathbb{E}\left[\left\lvert\hat{s}^{(i)}[q]-s[q]\right\rvert^2\right].
\end{equation}
Here we assume symbol estimation errors $\Delta s^{(i)}[q]$ are independently distributed and have zero means. Since the transmitted symbol $s[q]$ also has zero mean, the estimator $\hat{s}^{(i)}[q]$ is unbiased due to linearity. As a sufficiently large $MN$ is used in our simulation, we assume the error variance equal to MSE
\begin{equation}\label{eq:sigma2=mse}
    \left(\sigma_e^2\right)^{(i)}=\mse_{s}^{(i)}.
\end{equation}
Hence, the practical per-iteration error variance is extracted from simulation data by (\ref{eq:mse}) and (\ref{eq:sigma2=mse}). Given a channel realization, $\left(\sigma_e^2\right)^{(i-1)}$ and $\left(\sigma_e^2\right)^{(i)}$ can be directly put into (\ref{eq:sinr_analytical}) and (\ref{eq:sinr_soft_analytical}), which yields the theoretical post-equalization SINR in the $i$-th iteration. 

\begin{figure}[ht]
    \centering
    \includegraphics[width=8.0cm,trim={18 4 34 18},clip]{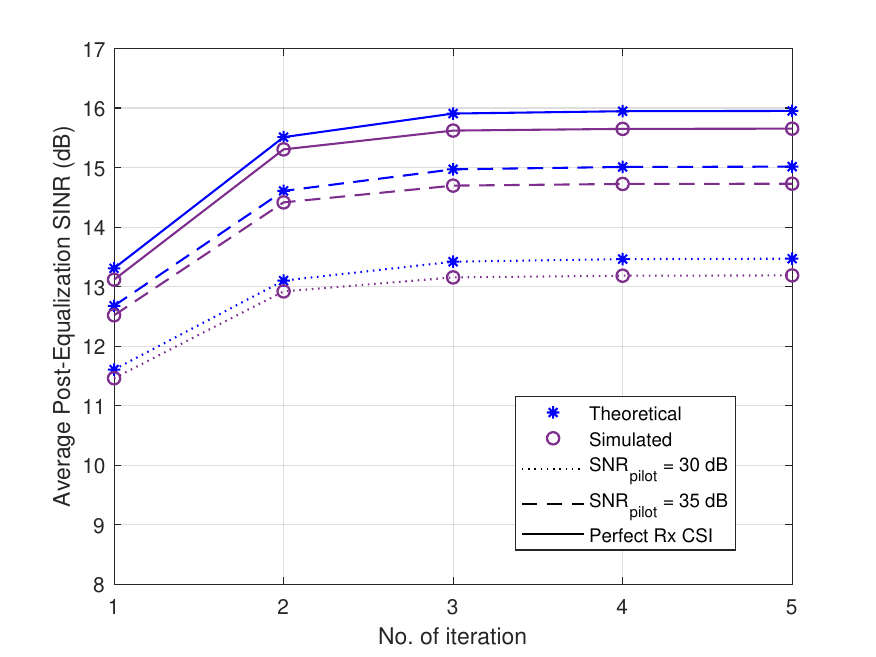}
    \caption{Simulated and theoretical SINR of MRC.}
    \label{fig:sinr_ite_mrc_16db}
\end{figure}

\begin{figure}[ht]
    \centering
    \includegraphics[width=8.0cm,trim={18 4 34 18},clip]{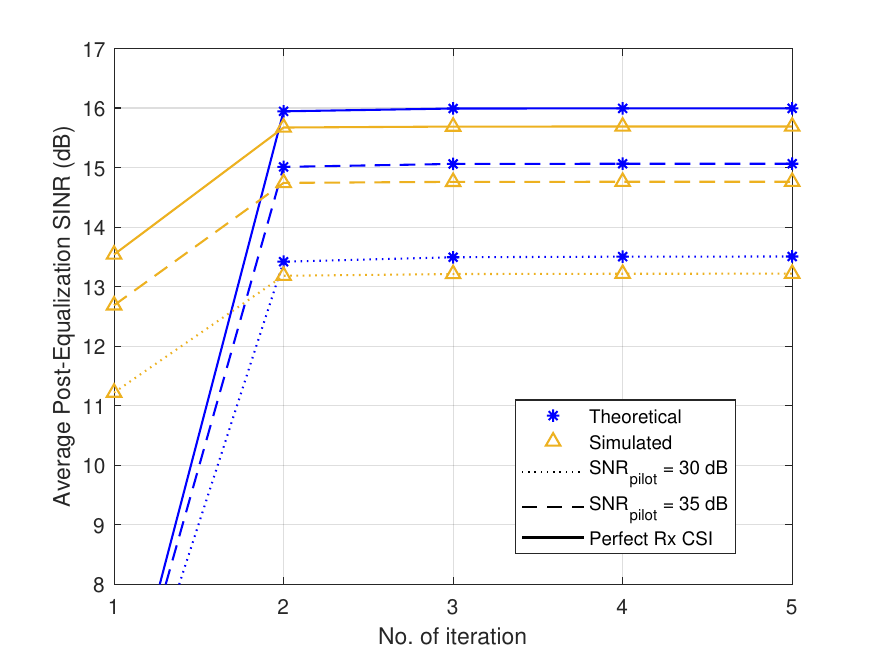}
    \caption{Simulated and theoretical SINR of hard SIC-MMSE.}
    \label{fig:sinr_ite_hardsicmmse_16db}
\end{figure}

\begin{figure}[ht]
    \centering
    \includegraphics[width=8.0cm,trim={18 4 34 18},clip]{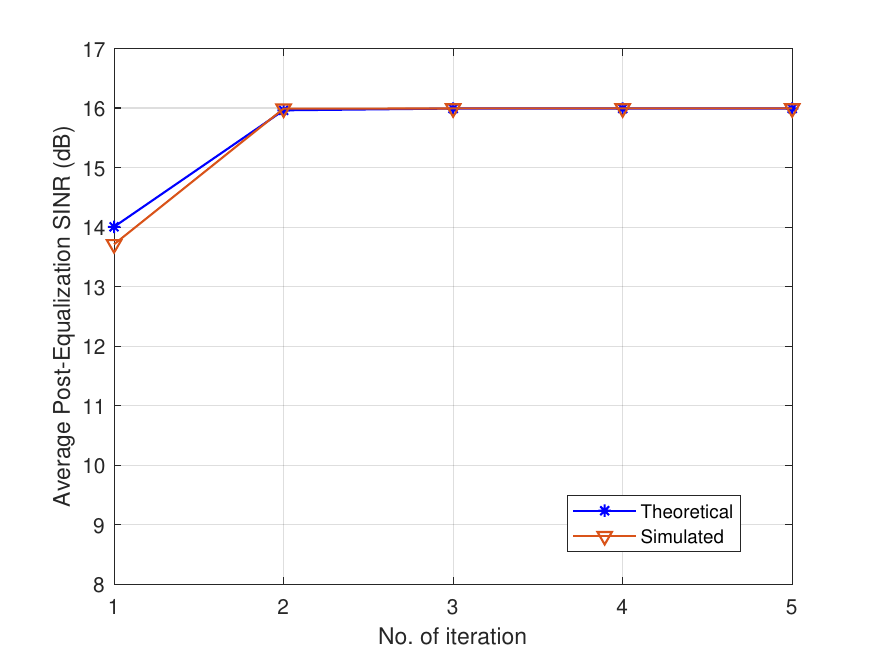}
    \caption{Simulated and theoretical SINR of soft SIC-MMSE.}
    \label{fig:sinr_ite_softsicmmse_16db}
\end{figure}

With an SNR of 16 dB, the derived post-equalization SINR is averaged over symbol index $q$ and plotted as a function of detection iteration number for MRC, hard SIC-MMSE, and soft SIC-MMSE, in Figs. \ref{fig:sinr_ite_mrc_16db}, \ref{fig:sinr_ite_hardsicmmse_16db}, and \ref{fig:sinr_ite_softsicmmse_16db}, respectively. We can see that, for both MRC and hard SIC-MMSE, the curves of theoretical SINR closely approach that of simulated SINR from the second iteration to the fifth iteration under different levels of channel estimation errors. Note that the theoretical SINR of the first iteration of hard SIC-MMSE departs from the practical value because hard SIC-MMSE has a distinct form in its first iteration, which is not considered in our analysis. For soft SIC-MMSE without channel estimation error, the derived SINR also matches the simulated SINR. This suggests that our analytical models for post-equalization SINR are robust across various error states. When the variances of symbol estimation errors and channel estimation errors are given, our equations provide a close approximation for the post-equalization SINR of MRC, hard SIC-MMSE, and soft SIC-MMSE. 

\subsection{SINR Upper Bound}\label{sec:sinr_upperbound}
With the analytical post-equalization SINR in Section \ref{sec:post_eq_sinr_i}, we now present the SINR upper bound for MRC and SIC-MMSE by assuming ideal interference cancellation. Note that the ideal interference cancellation is used only to find the performance benchmark. Regarding Proposition \ref{proposition:sinr_hard=sinr_mrc}, for the second and following iterations, hard SIC-MMSE and MRC have identical theoretical SINR given the same symbol estimation errors and channel estimation errors. Furthermore, it can be seen from (\ref{eq:softmmse_filter}) that the soft SIC-MMSE filter becomes equivalent to the hard SIC-MMSE filter when there is no ISI. Therefore, with ideal interference cancellation, MRC, hard, and soft SIC-MMSE, all share the same theoretical SINR defined in Proposition \ref{proposition:sinr_mrc}. The corresponding SINR upper bound is obtained by substituting $\sigma_e^2=0$ to (\ref{eq:sinr_analytical}).

\begin{figure}[ht]
    \centering
    \includegraphics[width=8.0cm,trim={18 4 34 18},clip]{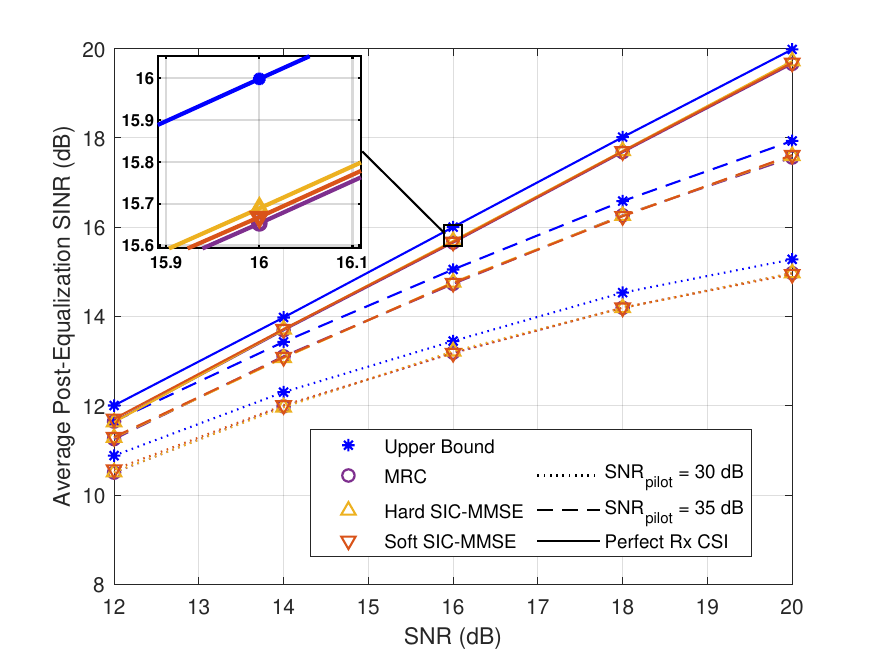}
    \caption{SINR upper bound and the simulated SINR of MRC, hard, and soft SIC-MMSE.}
    \label{fig:sinr_upperbound}
\end{figure}

A comparison between the SINR upper bound and the simulated SINR is plotted in Fig. \ref{fig:sinr_upperbound}. Across the plotted SNR region, our SINR upper bound remains tight under various channel estimation errors. An obvious reduction in slope can be observed under high channel estimation errors. This visualizes the performance degradation of these linear detectors caused by imperfect CSI. In addition, both analytical and numerical results indicate that MRC, hard, and soft SIC-MMSE share similar SINR performance, closely approaching the upper bound assuming perfect interference cancellation. This observation suggests that these detectors show comparable capabilities in reducing the residual interference power to a nearly optimal level.

\section{BER Analysis and Improved MRC Detectors}\label{sec:ber&improve}

In this section, we analyze the BER performance of MRC and SIC-MMSE. We highlight the limitation of MRC in fully canceling residual interference and investigate its error floor from the perspective of correlation propagation. Based on this BER analysis, two improved MRC detectors, MRC-SD and SSMI-MRC, are proposed to enhance the BER performance of MRC.


\subsection{BER Analysis}\label{sec:ber_gaussian}

Using the analytical equations for post-equalization SINR in Section \ref{sec:post_eq_sinr_i}, we employ state evolution to approximate the BER of MRC and SIC-MMSE at their convergence. Starting from the first iteration, the theoretical post-equalization SINR for the $q$-th symbol in the $i$-th iteration $\mathrm{SINR}_q^{(i)}$ is computed using (\ref{eq:sinr_analytical}) for MRC and hard SIC-MMSE, or (\ref{eq:sinr_soft_analytical}) for soft SIC-MMSE. With theoretical SINR, we want to find the post-ML MSE to serve as the variance of symbol estimation errors in the next iteration following (\ref{eq:sigma2=mse}). However, the accurate computation of MSE requires full knowledge of the exact distribution of the post-equalization noise plus interference component $\eta^{(i)}$ in (\ref{eq:s_tilde_mrc}), (\ref{eq:s_tilde_hardsicmmse}), or (\ref{eq:s_tilde_softsicmmse}), which is typically unavailable due to the non-linear ML detection across iterations. Therefore, we assume the post-equalization noise plus interference component $\eta^{(i)}$ follows a complex Gaussian distribution. Then, the post-ML symbol error rate (SER) can be approximated by the union bound of the AWGN channel
\begin{equation}\label{eq:unionbound}
    \mathrm{SER}^{(i)} \leq (A-1)Q\left(\sqrt{\mathbb{E}_q\left[\mathrm{SINR}_q^{(i)}\right]\frac{d_\mathrm{min}^2(\Lambda)}{2P_t}}\right),
\end{equation}
where $d_{\min}(\Lambda)$ is the minimum distance of the alphabets $\Lambda$. Using the SER upper bound and only assuming adjacent symbol errors, we can approximate the post-ML MSE and BER by
\begin{equation}\label{eq:ser2mse}
    \mathrm{MSE}_s^{(i)} \approx d_\mathrm{min}^2(\Lambda)\mathrm{SER}^{(i)},
\end{equation}
and
\begin{equation}\label{eq:ser2ber}
    \mathrm{BER}^{(i)} \approx \frac{\mathrm{SER}^{(i)}}{\log_2\abs{\Lambda}}.
\end{equation}

Referring to (\ref{eq:sigma2=mse}), the post-ML MSE approximated by (\ref{eq:ser2mse}) can be directly used as the updated variance of symbol estimation errors $\left(\sigma_e^2\right)^{(i)}$ to compute $\mathrm{SINR}_q^{(i+1)}$ in the next iteration. At convergence, the symbol estimates yield a steady-state MSE, i.e., $\left(\sigma_e^2\right)^{(i)} = \left(\sigma_e^2\right)^{(i-1)}$ as $i\rightarrow\infty$. The converged BER is then approximated by evaluating (\ref{eq:ser2ber}) at the error state characterized by $\bigl.\left(\sigma_e^2\right)^{(i)}\bigr|_{i\rightarrow\infty}$. To ensure full convergence, the state evolution is run for 20 iterations.

\begin{figure}[ht]
    \centering
    \includegraphics[width=8.0cm,trim={18 4 34 18},clip]{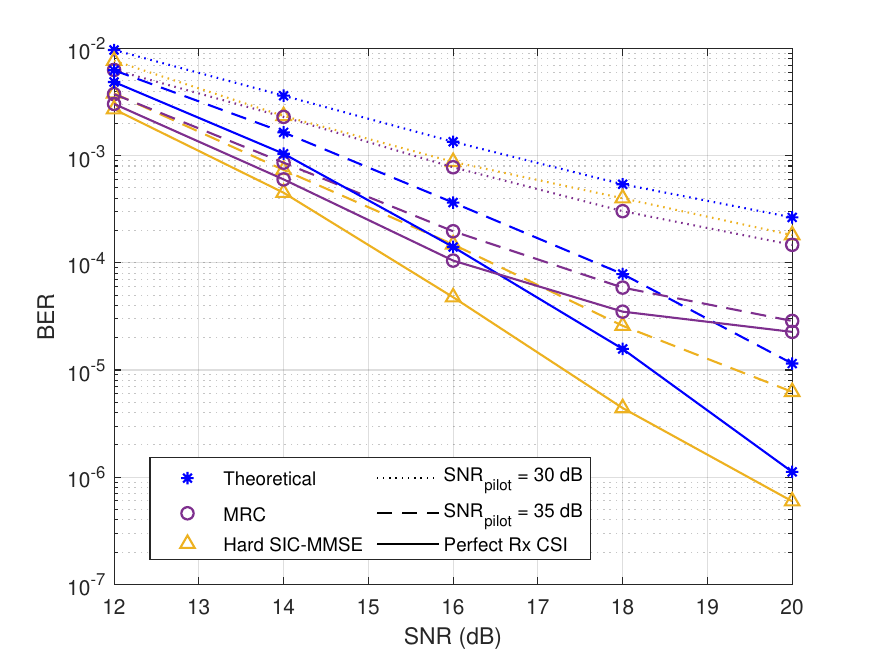}
    \caption{Simulated and theoretical BER of MRC and hard SIC-MMSE by state evolution.}
    \label{fig:ber_csi_analysis_hard_mrc}
\end{figure}

\begin{figure}[ht]
    \centering
    \includegraphics[width=8.0cm,trim={18 4 34 18},clip]{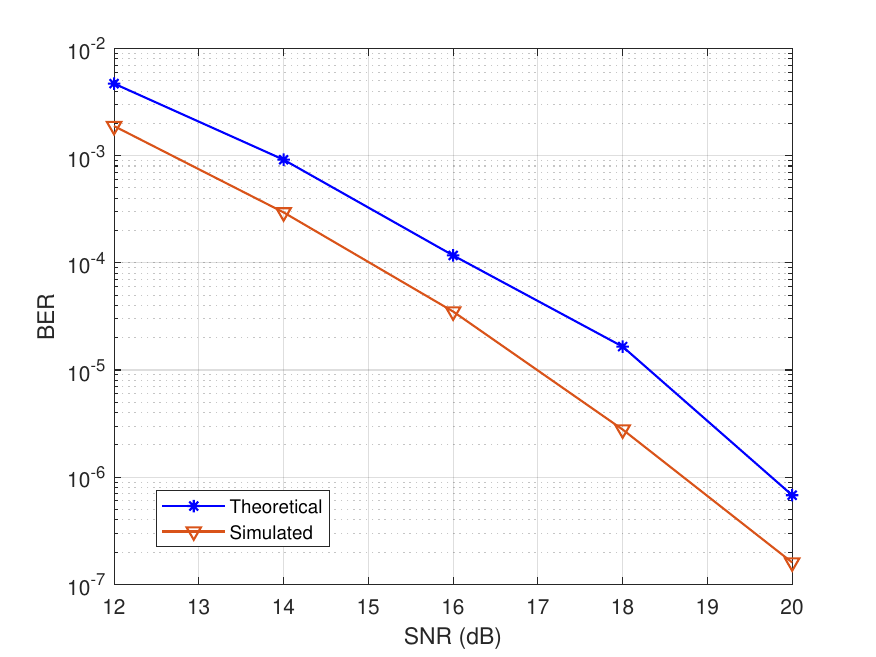}
    \caption{Simulated and theoretical BER of soft SIC-MMSE by state evolution.}
    \label{fig:ber_csi_analysis_softsicmmse}
\end{figure}

Because the first iteration of hard SIC-MMSE is not considered in Proposition \ref{proposition:sinr_hard=sinr_mrc}, the BER approximation by state evolution leads to identical results for MRC and hard SIC-MMSE. The theoretical BER performance of hard SIC-MMSE and MRC in the presence of channel estimation errors is presented in Fig. \ref{fig:ber_csi_analysis_hard_mrc}. The theoretical BER performance of soft SIC-MMSE without channel estimation error is presented in Fig. \ref{fig:ber_csi_analysis_softsicmmse}. The corresponding simulated BER performance of these detectors from Section \ref{sec:numerical_results} is also plotted for comparison. 

For both hard and soft SIC-MMSE detectors, the analytical BER by state evolution provides an effective approximation of the BER performance across all simulated SNR levels and channel estimation error conditions. However, for the MRC detector, the simulated BER exhibits an error floor that deviates from the theoretical BER curve. This is because hard-decision interference cancellation is susceptible to non-Gaussian residual interference \cite{Delic2002RobustCDMA}. As such intractable residual interference becomes more prominent in the high SNR region, an error floor emerges. Therefore, the error floor cannot be well captured under the Gaussian noise assumption.

Despite the limitation, the analytical BER for MRC remains accurate under high channel estimation errors or low SNR, where residual interference is effectively masked by Gaussian noise. This BER analysis thus allows accurate evaluation of the impact of imperfect CSI on the performance of MRC and SIC-MMSE without computationally heavy Monte Carlo simulations.

\subsection{Error Floor of MRC}\label{sec:err_floor}

We now provide a detailed analysis for the error floor of iterative MRC shown in Fig. \ref{fig:ber_csi_analysis_hard_mrc}. Our investigation focuses on the propagation of correlation between consecutive symbol estimates, revealing the limitation of hard-decision interference cancellation in eliminating non-Gaussian residual interference. To see this, consider the MRC output in (\ref{eq:s_tilde_mrc}). Without loss of generality, we consider perfect CSI and drop the iteration index for simplicity. Then, the MRC output gives
\begin{align}\label{eq:mrc_perfcsi}
    \tilde{s}[q] &= s[q]+(\mathbf{g}_q^\herm/v_q^\mathrm{MRC})\tilde{\mathbf{z}}_q,
\end{align}
with RIPN
\begin{align}
    \tilde{\mathbf{z}}_q = \mathbf{z}_q &-\sum_{\mathclap{\Delta l\neq 0,\Delta l\in\dot{\mathcal{L}}}} \mathbf{g}_{q,\Delta l}\Delta s\bigl[[q+\Delta l]_{MN}\bigr].
\end{align}
Due to residual interference, $\tilde{s}[q]$ has a non-zero covariance with its consecutive symbol estimates $\tilde{s}[[q+\Delta l]_{MN}],\Delta l\in\dot{\mathcal{L}}$, denoted as
\begin{align}\label{eq:cov_stilde}
    C_{q,[q+\Delta l]_{MN}} = -\frac{\mathbf{g}_q^\herm\mathbf{g}_{q,\Delta l}}{v_q^\mathrm{MRC}} \var(\Delta\tilde{s}[q+\Delta l]).
\end{align}
It can be seen that MRC filtering does not suppress the correlation between consecutive symbol estimates. Therefore, in the MRC detector, decorrelation relies only on interference cancellation and the gain from element-wise ML detection.

After domain conversion by (\ref{eq:dft_sm2xm}), we denote the DD domain post-equalization symbol error as
\begin{align}\label{eq:xtilde_deltaxtilde}
    \Delta\tilde{x}_m[n] = \tilde{x}_m[n] - x_m[n],
\end{align}
which is also the error at the input of the ML detector. Because the DFT matrix is unitary, the correlation between consecutive symbol estimates is preserved. Thereby, similar to (\ref{eq:vardx=vards}), the DD domain covariance can be approximated by
\begin{align}\label{eq:cov_xtilde}
    \cov(\tilde{x}_m[n],\tilde{x}_{m+\Delta l}[n]) = \mathbb{E}_{\dot{n}}\mkern-3mu\left[C_{\dot{n}M+m,[\dot{n}M+m+\Delta l]_{MN}}\right]
\end{align}
for $\Delta l\in\dot{\mathcal{L}}$.

When the prior symbol estimates used for interference cancellation in (\ref{eq:mrc_perfcsi}) are not i.i.d., as indicated by (\ref{eq:cov_stilde}) and (\ref{eq:cov_xtilde}), the correlation between symbol estimates persists after MRC equalization and domain conversion. Moreover, $\Delta\tilde{x}_m[n]$ cannot be modeled as Gaussian since the central limit theorem no longer applies. Instead, $\Delta\tilde{x}_m[n]$ behaves like an impulse noise \cite{Delic2002RobustCDMA}. Notably, such impulse noises are sporadic, so the impact of the loss of Gaussianity is not captured by the SINR analysis in Section \ref{sec:sinr_analysis} and the BER analysis in Section \ref{sec:ber_gaussian}. Since the ML detection in (\ref{eq:ml}) is not robust to impulse errors, the hard-decision symbol estimates $\hat{x}_m[n]$ contains occasional outliers, which carries forward the correlation. Consequently, correlation and residual interference can persist across iterations.

In general, the error floor of iterative MRC is caused by the limitation of hard-decision interference cancellation in fully decorrelating consecutive symbol estimates, which manifests as non-Gaussian residual interference at the input of the ML detector. This also highlights the possibility of BER improvement for MRC, which is explored in the following section.

\subsection{MRC with Subtractive Dither}\label{sec:mrc_dither}

Based on our previous analysis, we now propose MRC with subtractive dither (MRC-SD) to achieve BER gains for MRC with a negligible increase in detection complexity. As stated in Section \ref{sec:err_floor}, the error floor of MRC stems from the persistent correlation between consecutive symbol estimates. Thereby, we want to whiten prior symbol estimation errors so that the Gaussianity of $\Delta\tilde{x}_m[n]$ can be recovered. As opposed to the covariance-aware MMSE equalization described in Section \ref{sec:softsicmmse}, the proposed dither-aided method indirectly deccorrelates symbol estimates by breaking correlation propagation.

It is known that subtractive dither can fully decorrelate the output and input of a quantizer \cite{Schuchman1964SubDither}. In iterative MRC, the discretization by the ML detection resembles a quantization process, allowing us to leverage the principle of subtractive dither to decorrelate the equalized symbols $\tilde{x}_m[n]$ and the hard-decision estimates $\hat{x}_m[n]$. Consider a subtractive dither signal $d_m[n]$, which is applied to the element-wise ML detector in (\ref{eq:ml}) as
\begin{align}\label{eq:sub_dither}
    \hat{x}_m[n] &= \argmin_{a_j\in\Lambda}{\abs{a_j-(\tilde{x}_m[n]+d_m[n])}}-d_m[n].
\end{align}
We define $d_m[n]$ to be independently uniformly distributed over $\mathcal{U}\triangleq\left\{u\in\mathbb{C}\middle|\Re(u),\Im(u)\in\left[-\delta_d,\delta_d\right]\right\}$ with a bounding constant $\delta_d\in(0,d_{\min}(\Lambda)/2)$, denoted as
\begin{align}\label{eq:d_uniform}
    d_m[n] \overset{\scalebox{0.5}{unif}}{\sim} \mathcal{U}.
\end{align}
As a necessary condition for (\ref{eq:d_uniform}), $d_m[n]$ is bounded in the complex plane as $\Re(d_m[n]),\Im(d_m[n])\in[-\delta_d,\delta_d]$.

Then, we express the input to the ML detector as
\begin{align}\label{eq:xtilde_deltadelta}
    \tilde{x}_m[n] = x_m[n] + \Delta\tilde{x}_m^d[n] + \Delta\tilde{x}_m^i[n],
\end{align}
where we define the fractional input error $\Delta\tilde{x}_m^d[n]$ bounded as $\Re\left(\Delta\tilde{x}_m^d[n]\right),\Im\left(\Delta\tilde{x}_m^d[n]\right)\in[-\delta_d,\delta_d]$, and the integer input error $\Delta\tilde{x}_m^i[n]\in 2\delta_d\mathbb{Z}[i]$. This ensures that the total input error $\Delta\tilde{x}_m^d[n] + \Delta\tilde{x}_m^i[n]$ can take any value in the complex plane. Similarly, based on (\ref{eq:sub_dither}), we write the output of the ML detector as
\begin{align}\label{eq:xhat_ddelta}
    \hat{x}_m[n] &= x_m[n] - d_m[n] + \Delta\hat{x}_m^i[n],
\end{align}
where $\Delta\hat{x}_m^i[n]\in d_{\min}(\Lambda)\mathbb{Z}[i]$ is the integer output error, with the fractional component determined by the dither $d_m[n]$.

\begin{figure}[ht]
    \centering
    \includegraphics[width=8.0cm,trim={10 0 15 18},clip]{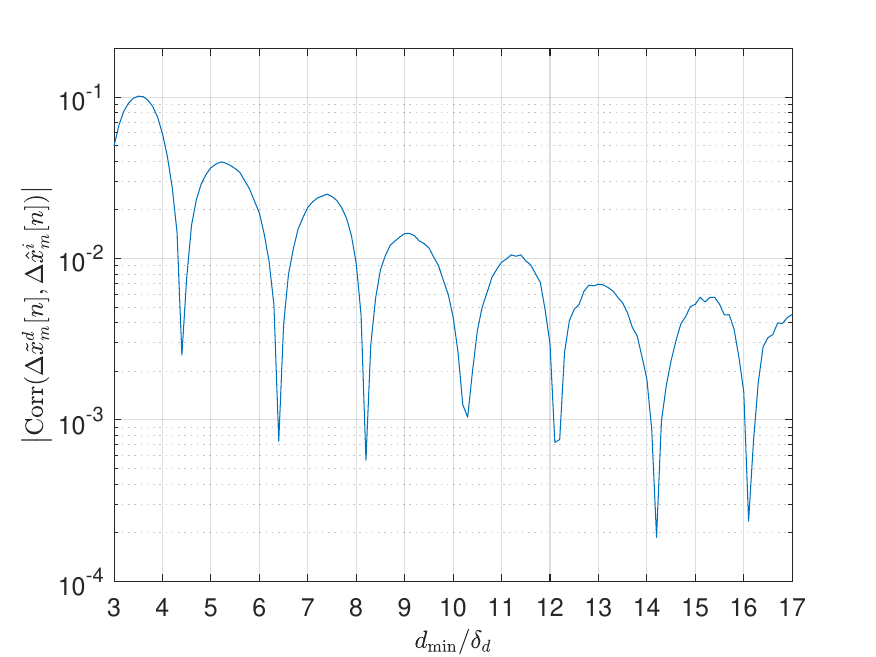}
    \caption{Correlation between $\Delta\hat{x}_m^i[n]$ and $\Delta\tilde{x}_m^d[n]$ for 4-QAM.}
    \label{fig:corr_io}
\end{figure}

The integer input error $\Delta\tilde{x}_m^i[n]$ forms a lattice in the complex plane with a resolution of $2\delta_d$. The integer output error $\Delta\hat{x}_m^i[n]$ forms another lattice with a resolution of $d_{\min}$. When the resolution of $\Delta\tilde{x}_m^i[n]$ increases, $\Delta\hat{x}_m^i[n]$ becomes less correlated with $\Delta\tilde{x}_m^d[n]$. To verify this, we plot the correlation between $\Delta\hat{x}_m^i[n]$ and $\Delta\tilde{x}_m^d[n]$ as a function of $d_{\min}(\Lambda)/\delta_d$ for 4-QAM in Fig. \ref{fig:corr_io}. We observe that $\abs{\mathrm{Corr}\left(\Delta\hat{x}_m^i[n],\Delta\tilde{x}_m^d[n]\right)}$ approaches zero as $d_{\min}(\Lambda)/\delta_d$ increases, implying that $\Delta\hat{x}_m^i[n]$ can be independent of $\Delta\tilde{x}_m^d[n]$ for $2\delta_d\ll d_{\min}(\Lambda)$. Then, for fractional input and output errors, $\Delta\tilde{x}_m^d[n]$ and $d_m[n]$, which share identical bounds in the complex plane by definition, the ML detector for rectangular QAM can be viewed as a 2D finite-level quantizer with uniformly spaced quantization levels $2\delta_d$ apart. And the unbounded integer output error $\Delta\hat{x}_m^i[n]$ is only dependent on the integer input error $\Delta\tilde{x}_m^i[n]$.

For a quantizer subtractively dithered by $d_m[n]$, the Schuchman's condition \cite{Schuchman1964SubDither} states that, when the characteristic function of $d_m[n]$ satisfies
\begin{align}\label{eq:schuchman_cond}
    P_d\left(\frac{\kappa}{2\delta_d}\right)=0,\forall \kappa\in \mathbb{Z}_{\ne 0},
\end{align}
the quantization error $\Delta\tilde{x}_m^d[n]$ with arbitray distribution can be rendered statistically independent of $d_m[n]$. In particular, $d_m[n]$ following a uniform distribution is the simplest case that satisfies Schuchman's condition in (\ref{eq:schuchman_cond}). Then, $\Delta\hat{x}_m^d[n]$ can be rendered independent of $d_m[n]$, i.e.,
\begin{align}
    p\left(\Delta\tilde{x}_m^d[n]\middle| d_m[n]\right) = p\left(\Delta\tilde{x}_m^d[n]\right).
\end{align}
That is, for the bounded error component $\Delta\tilde{x}_m^d[n]$, its correlation with consecutive symbol estimates is not transferred to the ML output and is therefore effectively eliminated. This allows the subsequent hard-decision interference cancellation to benefit from the enhanced i.i.d. property, nudging symbol estimates towards a state with reduced residual interference.

Despite the decorrelation effect for bounded errors, the subtractive dither $d_m[n]$ also introduces a noise source to MRC-SD. As shown in (\ref{eq:xhat_ddelta}), $d_m[n]$ behaves like a constant symbol estimation error for the ML output $\hat{x}_m[n]$, even in the absence of integer output error $\Delta\hat{x}_m^i[n]$. For $d_m[n]$ defined in (\ref{eq:d_uniform}), its variance is given by $\sigma_d^2=\frac{1}{3}\delta_d^2$. By substituting $\sigma_d^2$ for $\sigma_e^2$ in (\ref{eq:sinr_analytical}) and ignoring the channel estimation error and AWGN, the post-equalization SINR of MRC-SD has an upper bound of
\begin{align}\label{eq:sinr_mrcsd}
    \mathrm{SINR}_q^{\mathrm{MRC-SD}} < \frac{P_t\bigl(\mathbf{g}_q^\herm\mathbf{g}_q\bigr)^2}{\epsilon_d^2+\sigma_z^2\mathbf{g}_q^\herm\mathbf{g}_q},
\end{align}
where
\begin{align}\nonumber
    \epsilon_d^2 = \sigma_d^2\sum_{\mathclap{l_1,l_2\in\hat{\mathcal{L}}}}g_q^*[l_1]g_q[l_2]\sum_{\mathclap{\Delta l\neq0,\Delta l\in\dot{\mathcal{L}}}}g_{q,\Delta l}[l_1]g_{q,\Delta l}^*[l_2]
\end{align}
captures the impact of $d_m[n]$. Apparently, an error floor can be expected for MRC-SD, with an asymptotic post-equalization SINR of $\frac{P_t}{\epsilon_d^2}\bigl(\mathbf{g}_q^\herm\mathbf{g}_q\bigr)^2$. This means $\delta_d$ should be carefully selected to strike a balance between the decorrelation capability and the dither-induced error. We perform simulations to incrementally search for an appropriate $\delta_d$. As presented in Section \ref{sec:numerical_results}, with the properly selected $\delta_d=d_{\min}(\Lambda)/9.4$, MRC-SD can effectively improve the BER of MRC and does not exhibit an obvious error floor.

Theoretically, the mitigation of non-Gaussian residual interference achieved by subtractive dither can benefit other hard-decision interference cancellation-based detectors, such as hard SIC-MMSE. However, as the performance loss from dither-induced error may outweigh the benefits, further investigation is needed to determine its applicability based on the level of non-Gaussian residual interference. This remains challenging due to the difficulty in capturing such interference through conventional SINR metrics, as discussed in Section \ref{sec:err_floor}. In addition, the optimal dither level has yet to be fully analyzed, requiring further exploration under varying SNR and CSI conditions in future studies.

For MRC-SD, the element-wise ML detector in (\ref{eq:ml}) of the original MRC detector is replaced by the element-wise ML detector with subtractive dither in (\ref{eq:sub_dither}). Since generating one dither sample $d_m[n]$ with a uniform distribution has a complexity order of $\mathcal{O}(1)$, the increase in detection complexity is negligible. Therefore, the overall complexity order of MRC-SD is identical to that of the original MRC, i.e., $\mathcal{O}(n_\mathrm{ite}MN(\log_2N+\abs{\mathcal{L}}))$.

\subsection{SIC-MMSE Initialized MRC}\label{sec:ssmi_mrc}

According to Proposition \ref{proposition:sinr_hard=sinr_mrc}, the MRC and hard SIC-MMSE detectors only differ by their initial symbol estimates. Therefore, the hard SIC-MMSE detector can be viewed as a hard SIC-MMSE initialized MRC (HSMI-MRC) detector. As discussed in Section \ref{sec:err_floor}, MRC suffers from an error floor due to the limitation of hard-decision interference cancellation in fully canceling non-Gaussian residual interference. However, despite employing the same hard-decision scheme, hard SIC-MMSE does not exhibit an error floor in Fig. \ref{fig:ber_csi_analysis_hard_mrc}. This suggests that initial estimates play an important role in the BER performance of detectors based on iterative hard-decision interference cancellation, supporting Remark \ref{remark:1st_ite}.

By leveraging MMSE filtering, the first iteration of hard SIC-MMSE significantly reduces the correlation between consecutive symbol estimates. This allows the Gaussianity of $\Delta\tilde{x}_m[n]$ to be well retained in subsequent iterations. Similarly, soft SIC-MMSE can be employed to initialize symbols with even lower correlations between consecutive symbol estimates, thereby facilitating the effectiveness of hard-decision interference cancellation. Specifically, we propose the soft SIC-MMSE initialized MRC (SSMI-MRC), where soft SIC-MMSE, as described in Section \ref{sec:softsicmmse}, is run for one iteration, followed by iterative MRC. 

Because the first iteration of hard SIC-MMSE has the same computational complexity as that of soft SIC-MMSE, the overall complexity of SSMI-MRC is identical to that of hard SIC-MMSE, i.e., $\mathcal{O}\left(n_\mathrm{ite}MN(\log_2N+\abs{\mathcal{L}})+MN(l_{\max}+1)^3\right)$.

\section{Numerical Results}\label{sec:numerical_results}

Here we present the BER performance for the ODDM system with practical channel estimation and compare different data detection algorithms. The simulation is performed by collecting more than 500 frame errors. To simulate the doubly-selective channel, we adopt the power delay profile in the EVA model and generate random Doppler shifts by Jakes's model \cite{Jakes1974MicrowaveCommunications} with a user equipment (UE) speed of 500 km/h. The other parameters used in the simulation are summarized in Table \ref{tab:param}. 

\begin{table}[ht]
    \centering
    \caption{Simulation parameters}
    \label{tab:param}
    \begin{tabular}{|c|c|c|c|c|}
        \hline Modulation Type & Carrier Frequency & $T$ & $M$ & $N$ \\
        \hline 4-QAM & 5 GHz & 66.67 \textmu s & 512 & 32 \\
        \hline
    \end{tabular}
\end{table}

Based on the aforementioned physical parameters and the assumption of an on-grid channel, the channel delay taps are derived as $l_p=[0,0,1,2,3,5,8,13,19]$, with a maximum delay index $l_{\max}=19$. The Doppler taps of the channel are randomly distributed in the range $k_p\in\{-5,\dots,5\}$ according to the Jakes's spectrum.

\subsection{Perfect Receiver CSI}\label{sec:numerical_perfcsi}
\begin{figure}[ht]
    \centering
    \includegraphics[width=8.0cm,trim={18 4 34 18},clip]{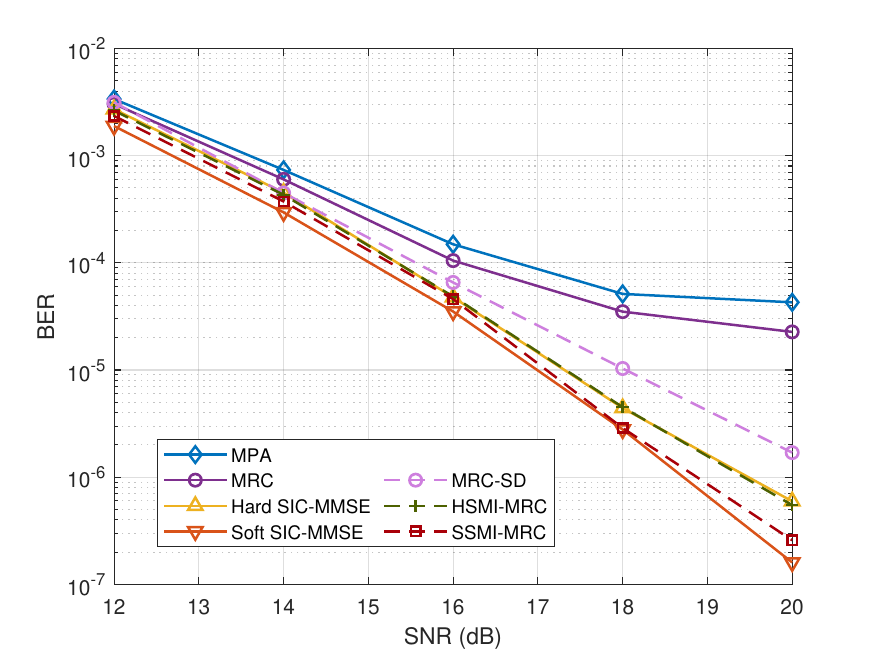}
    \caption{BER performance of ODDM with MPA, MRC, hard SIC-MMSE, soft SIC-MMSE, MRC-SD, HSMI-MRC, and SSMI-MRC under perfect Rx CSI.}
    \label{fig:ber_perfcsi}
\end{figure}

We first present the BER performance of detectors for ODDM with perfect CSI in Fig. \ref{fig:ber_perfcsi}, including the three modified MRC detectors, namely MRC-SD, HSMI-MRC, and SSMI-MRC. For all the algorithms, we set the maximum number of iterations to 10 to make sure the detection results have fully converged. For MRC-SD, we select a dither level of $\delta_d = d_{\min}(\Lambda)/9.4$ to achieve significant improvement in BER performance.

We observe that hard SIC-MMSE and soft SIC-MMSE have more steeply descending BER curves and they consistently outperform MPA and MRC across the entire SNR range. Moreover, soft SIC-MMSE achieves a performance gain over hard SIC-MMSE. On the other hand, both MPA and MRC exhibit error floors, whereas both hard and soft SIC-MMSE detectors show no sign of error floors within the SNR region of interest. This suggests that SIC-MMSE is more effective in interference cancellation and utilizing diversity.

For the MPA detector, the significant error floor indicates the existence of intractable residual interference. This is because the MPA detector assumes the Gaussianity of the interference plus noise term, which relies on the i.i.d. assumption of the interference components. However, as the algorithm iterates, symbol estimates become correlated, making the Gaussian assumption fail. Therefore, erroneous probability propagation arises when interference becomes dominant in the high SNR region \cite{Zhang2021AClipping}. In addition, the loopy belief propagation does not guarantee the convergence of MPA to an optimal fixed point.

Similarly, the error floor of MRC stems from residual interference that cannot be effectively mitigated by hard-decision interference cancellation, which has been discussed in Section \ref{sec:err_floor}. With the dither-aided approach detailed in Section {\ref{sec:mrc_dither}}, MRC-SD effectively lowers the error floor. Compared to MRC, MRC-SD provides significant improvement in BER while maintaining low detection complexity.

As implied by Proposition \ref{proposition:sinr_hard=sinr_mrc} and discussed in Section \ref{sec:ssmi_mrc}, the BER performance of MRC is heavily influenced by the accuracy of the initial symbol estimates. As shown in Fig. \ref{fig:ber_csi_all}, when the first iteration of hard SIC-MMSE is used to initialize symbol estimates for MRC, referred to as HSMI-MRC, the BER performance matches that of hard SIC-MMSE, confirming Proposition \ref{proposition:sinr_hard=sinr_mrc}. When soft SIC-MMSE is employed for symbol initialization, referred to as SSMI-MRC, even better BER performance is achieved. Furthermore, although SSMI-BER offers a BER gain over hard SIC-MMSE, it does so without increasing detection complexity order.

\subsection{Convergence}\label{sec:convergence}
\begin{figure}[ht]
  \centering
  \includegraphics[width=8.0cm,trim={18 4 34 18},clip]{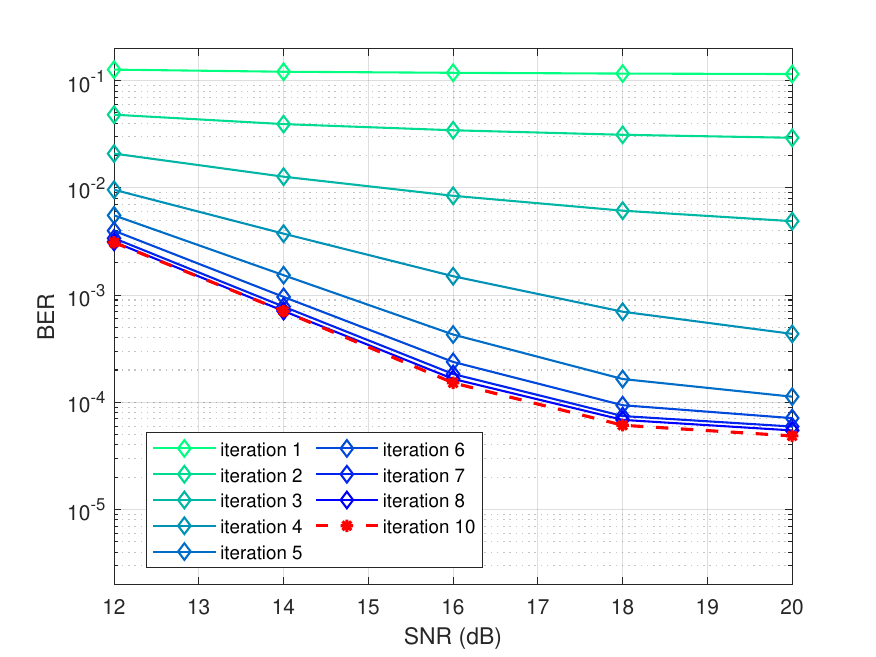}
  \caption{BER convergence of MPA detector.}
  \label{fig:conv_mpa}
\end{figure}

\begin{figure}[ht]
  \centering
  \includegraphics[width=8.0cm,trim={18 4 34 18},clip]{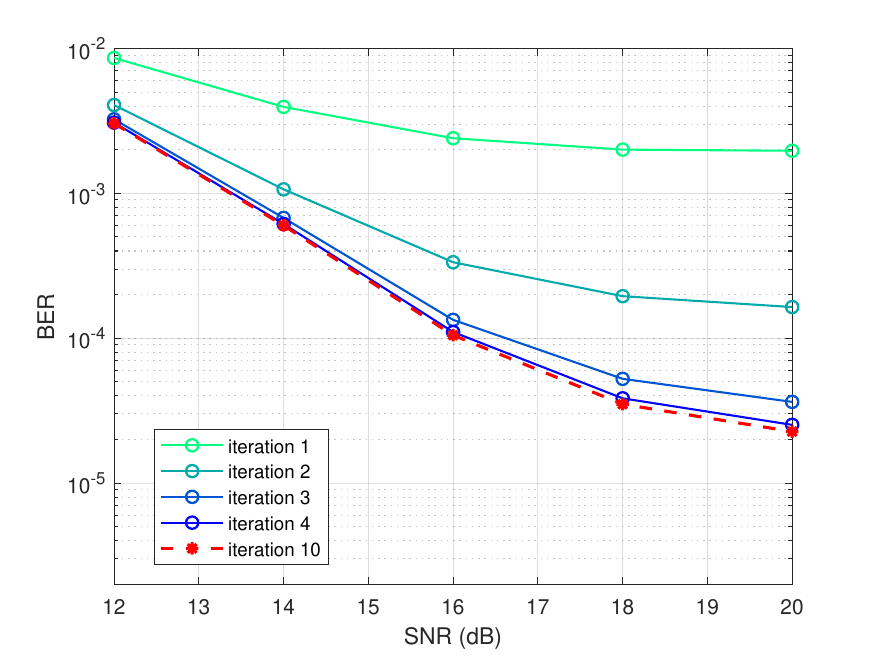}
  \caption{BER convergence of MRC detector.}
  \label{fig:conv_mrc}
\end{figure}

\begin{figure}[ht]
  \centering
  \includegraphics[width=8.0cm,trim={18 4 34 18},clip]{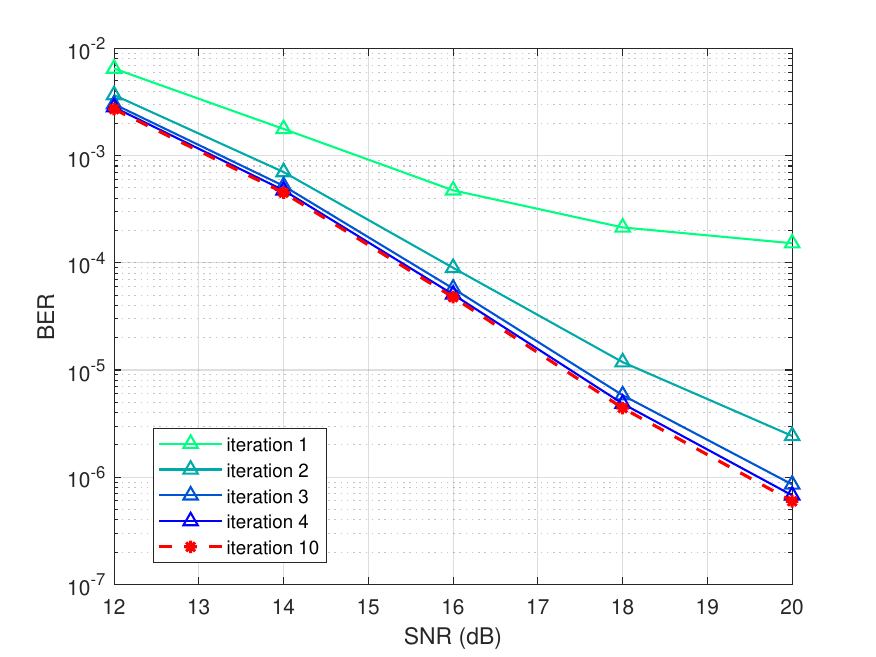}
  \caption{BER convergence of hard SIC-MMSE detector.}
  \label{fig:conv_hardsicmmse}
\end{figure}

\begin{figure}[ht]
    \centering
    \includegraphics[width=8.0cm,trim={18 4 34 18},clip]{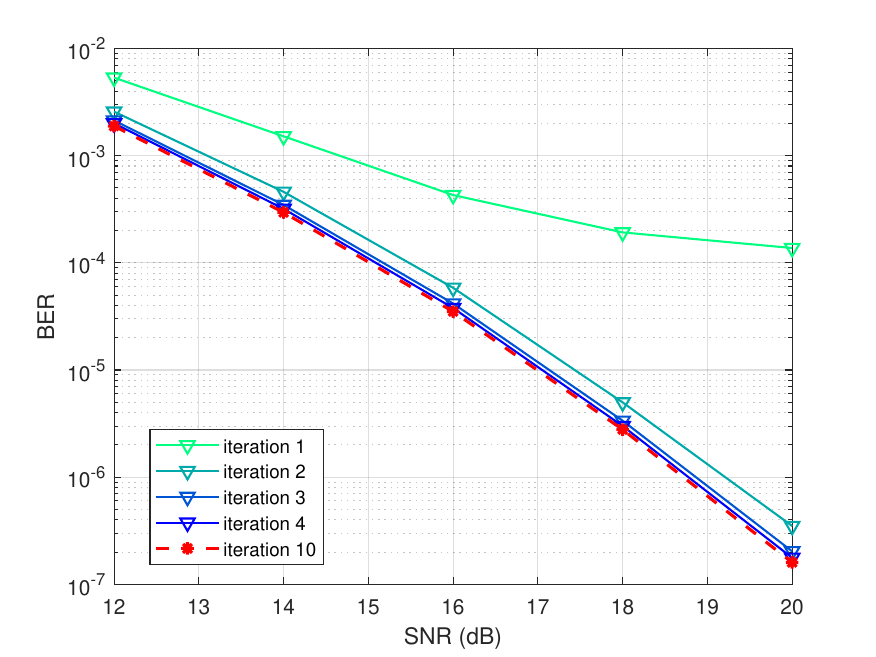}
    \caption{BER convergence of soft SIC-MMSE detector.}
    \label{fig:conv_softsicmmse}
\end{figure}

The convergence performance for MPA, MRC, hard SIC-MMSE, and soft SIC-MMSE are shown in Figs. \ref{fig:conv_mpa}, \ref{fig:conv_mrc}, \ref{fig:conv_hardsicmmse}, and \ref{fig:conv_softsicmmse}, respectively. We observe that MPA requires more iterations to converge compared to the other three detectors. Both the MRC and SIC-MMSE detectors converge after approximately 4 iterations. The short convergence cycle and reduced complexity of MRC make it the most computationally efficient. However, the matrix inversion required by SIC-MMSE introduces more complexity, making its overall complexity comparable to MPA.

\begin{figure}[ht]
    \centering
    \includegraphics[width=8.0cm,trim={18 4 34 18},clip]{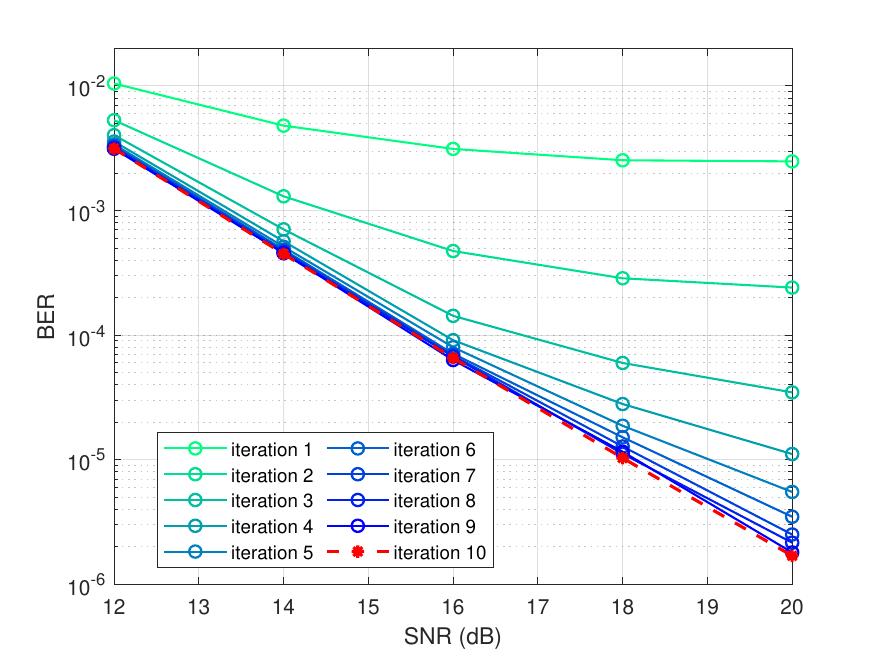}
    \caption{BER convergence of MRC-SD detector.}
    \label{fig:conv_mrcsd}
\end{figure}

The convergence performance of MRC-SD is presented in Fig. \ref{fig:conv_mrcsd}, highlighting the effect of subtractive dither. During the first two iterations, MRC-SD exhibits a higher BER than the original MRC. This is because hard-decision interference cancellation is more efficient in decorrelation before reaching the error floor, and the dither signal acts as a pure noise source. However, starting from the third iteration, MRC-SD outperforms MRC due to the additional decorrelation effect introduced by the subtractive dither. Despite overcoming the error floor of hard-decision interference cancellation, MRC-SD requires about 10 iterations to fully converge, which is significantly slower than hard SIC-MMSE and SSMI-MRC. This slower convergence is attributed not only to poorer initial estimates but also to the fact that the decorrelation effect of subtractive dither is limited to a bounded region, as discussed in Section \ref{sec:mrc_dither}, leading to incremental BER improvements across iterations.

\begin{figure}[ht]
    \centering
    \includegraphics[width=8.0cm,trim={18 4 34 18},clip]{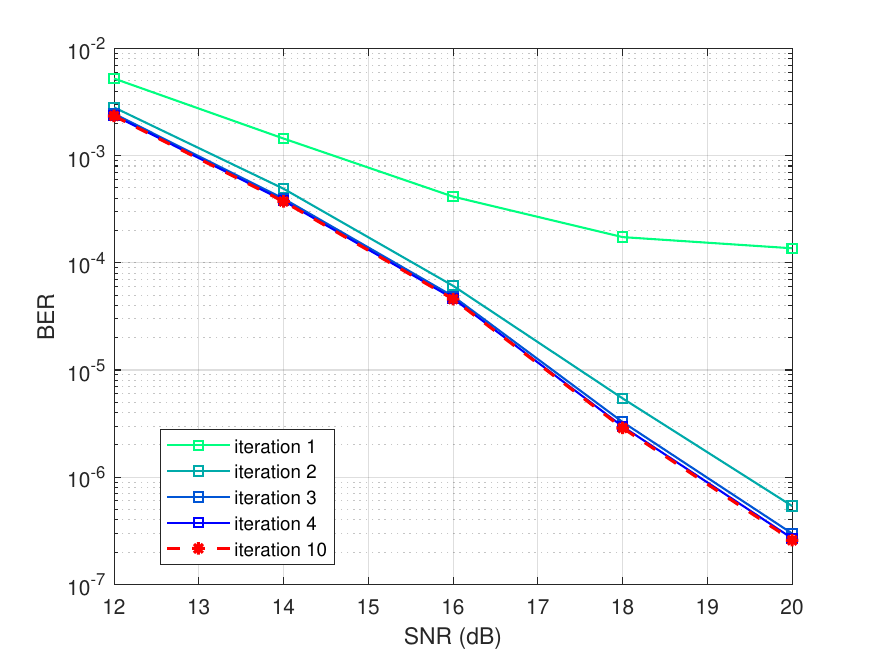}
    \caption{BER convergence of SSMI-MRC detector.}
    \label{fig:conv_ssmimrc}
\end{figure}

The convergence performance of SSMI-MRC is shown in Figs. \ref{fig:conv_ssmimrc}. Since soft SIC-MMSE is used to initialize the symbol estimates, the first iteration BER of SSMI-MRC is slightly lower than that of hard SIC-MMSE. The advantage of more accurate initial symbol estimates is enhanced in subsequent iterations, giving SSMI-MRC better overall BER performance. This underscores the significance of initial estimates on the performance of hard-decision interference cancellation-based detectors.

\subsection{Imperfect Receiver CSI}\label{sec:imperfect_csi}
\begin{figure}[ht]
    \centering
    \subfloat[SNR\textsubscript{pilot} = 30 dB]{\includegraphics[width=0.78\columnwidth,trim={18 4 18 23},clip]{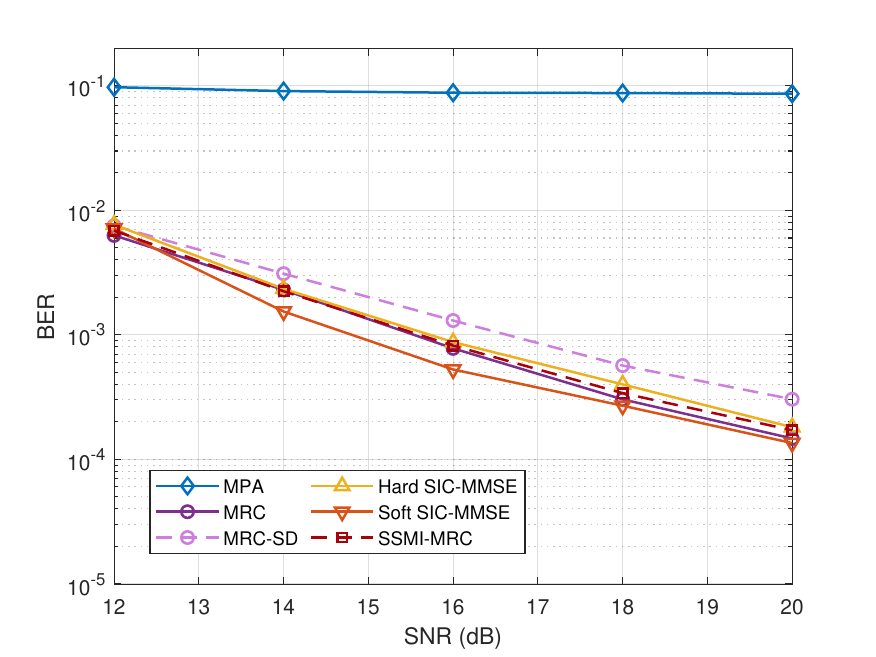}}\\
    \subfloat[SNR\textsubscript{pilot} = 35 dB]{\includegraphics[width=0.78\columnwidth,trim={18 4 18 18},clip]{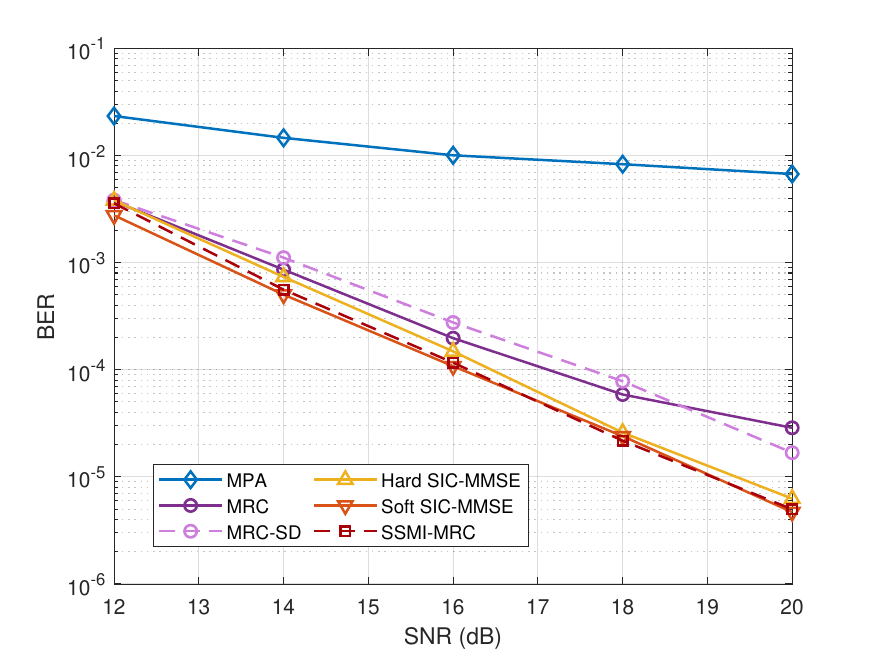}}\\
    \subfloat[SNR\textsubscript{pilot} = 40 dB]{\includegraphics[width=0.78\columnwidth,trim={18 4 18 18},clip]{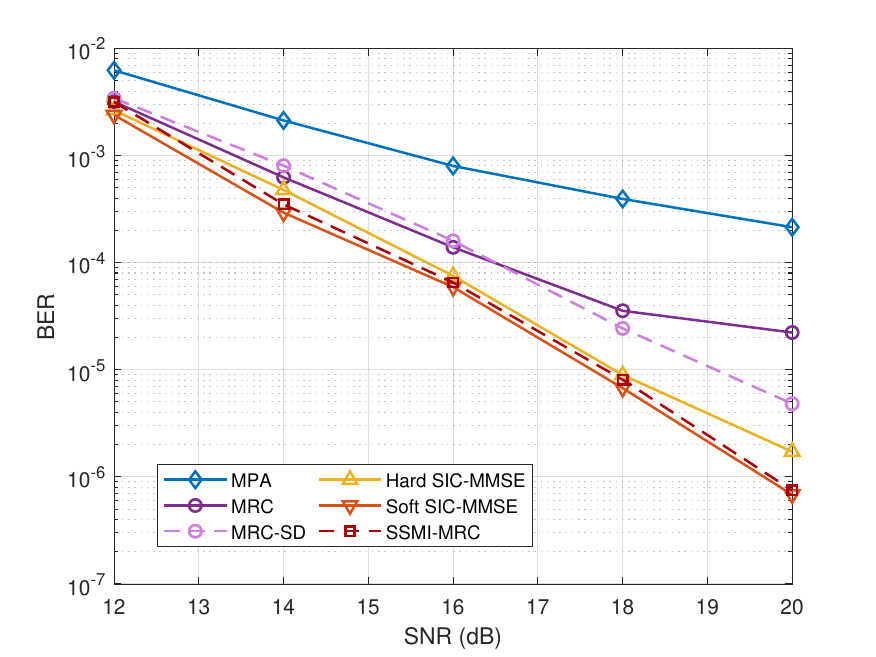}}\\
    \caption{BER performance of different detectors with imperfect receiver CSI.}
    \label{fig:ber_csi_all}
\end{figure}

As a final point, we examine the BER performance of the detectors with practical channel estimates. We obtain the channel estimates using the pilot-aided channel estimation scheme described in Section \ref{sec:chan_est}. The pilot SNR varies from 30 dB to 40 dB, indicating the channel estimation error from small to large. The results for MPA, MRC, MRC-SD, hard SIC-MMSE, soft SIC-MMSE and SSMI-MRC are shown in Fig. \ref{fig:ber_csi_all}. As channel estimation error increases, the BER gap between MRC, MRC-SD, hard SIC-MMSE, soft SIC-MMSE and SSMI-MRC becomes less distinct. The BER performance of MPA, however, has deteriorated much more significantly. This suggests that MPA is more sensitive to channel estimation errors.

Similar to MPA, MRC-SD also shows more significant performance degradation under imperfect CSI. Specifically, with a pilot SNR of 40 dB, MRC-SD is outperformed by MRC for $\text{SNR}<16\mathrm{\ dB}$. As the pilot SNR decreases further to 35 dB and 30 dB, MRC-SD progressively loses its advantage across all SNR levels. This degradation occurs because the use of subtractive dither introduces a noise source to MRC-SD, as shown in (\ref{eq:sinr_mrcsd}). Consequently, MRC-SD is outperformed by the original MRC detector when the performance loss due to dither-induced noise outweighs its benefits as a decorrelator. This is particularly evident under high channel estimation errors, where detection accuracy is limited by Gaussian noise rather than non-Gaussian residual interference.

\subsection{Discussion}\label{sec:discussion}

Our results showcase the BER performance of ODDM detectors with both perfect and imperfect CSI. Considering the parameters used in our simulations, we outline the representative BER results of these detectors in Table \ref{tab:ber}. When considering imperfect CSI with $\text{SNR}_\text{pilot}=40\mathrm{\ dB}$, the minimum SNR values required for MPA, MRC-SD, MRC, hard SIC-MMSE, SSMI-MRC, and soft SIC-MMSE to achieve a BER of $10^{-3}$ follow a decreasing order. The same order holds for MRC-SD, hard SIC-MMSE, SSMI-MRC, and soft SIC-MMSE to achieve a BER of $10^{-5}$ with perfect Rx CSI. However, MPA and MRC cannot reach that BER value due to their limitation in fully canceling residual interference, especially in high SNR scenarios.

\setlength{\tabcolsep}{0.5ex}
\begin{table}[ht]
    \centering
    \caption{BER Performance of ODDM Detectors}
    \label{tab:ber}
    \begin{tabular}{lcccccc}
        \toprule
        \multicolumn{1}{l}{\multirow{2}{6em}[-0.3em]{SNR (dB) \\required for}} & \multirow{2}{*}[-0.25em]{MPA} & \multirow{2}{2.5em}[-0.25em]{MRC} & \multicolumn{2}{c}{MRC w/} & \multicolumn{2}{c}{SIC-MMSE}\\
        \cmidrule{4-7}
        {} & {} & {} & \multirow{1}{1.5em}{SD} & \multirow{1}{3em}{SSMI} & \multirow{1}{2em}{hard} & \multirow{1}{2em}{soft}\\
        \midrule
        \multirow{1}{14em}{BER $=10^{-3}$, SNR\textsubscript{pilot} = 40 dB} & 15.5 & 13.4 & {13.7} & {13.0} & 13.1 & 12.8\\
        \midrule
        \multirow{1}{14em}{BER $=10^{-5}$, Perfect Rx CSI} & - & - & {18.0} & {17.1} & 17.3 & 17.0\\
        \bottomrule
    \end{tabular}
\end{table}

Following the discussion in Section \ref{sec:detectors} and \ref{sec:ber&improve}, we also summarize the complexity orders of the detectors in Table \ref{tab:complexity}, expressed as functions of frame size, number of paths, maximum delay spread, and number of iterations. By Table \ref{tab:ber} and \ref{tab:complexity}, we conclude that when low detection complexity is prioritized, the choice can be made between MRC-SD and MRC depending on the level of channel estimation errors. MRC-SD is preferred when accurate CSI is available. If detection accuracy is prioritized, soft SIC-MMSE offers superior BER performance across all SNR levels and CSI conditions. To strike a good balance between detection accuracy and complexity, SSMI-MRC should always be chosen over hard SIC-MMSE, as it delivers better BER performance without additional complexity. Therefore, MRC, MRC-SD, SSMI-MRC, and soft SIC-MMSE provide effective trade-offs between BER performance and detection complexity, which facilitates the implementation of ODDM modulation to satisfy practical system requirements.

\setlength{\tabcolsep}{0.5ex}
\begin{table}[ht]
    \centering
    \caption{Complexity of ODDM Detectors}
    \label{tab:complexity}
    \begin{tabular}{ll}
        \toprule
        {Detector} & \multicolumn{1}{l}{Complexity Order}\\
        \midrule
        MPA & {$\mathcal{O}(n_\mathrm{ite}MNP\abs{\Lambda})$}\\
        \midrule
        MRC & {$\mathcal{O}(n_\mathrm{ite}MN(\log_2N+\abs{\mathcal{L}}))$}\\
        \midrule
        MRC-SD & {$\mathcal{O}(n_\mathrm{ite}MN(\log_2N+\abs{\mathcal{L}}))$}\\
        \midrule
        SSMI-MRC & {$\mathcal{O}\left(n_\mathrm{ite}MN(\log_2N+\abs{\mathcal{L}})+MN(l_{\max}+1)^3\right)$}\\
        \midrule
        hard SIC-MMSE\hspace{0.4cm} & {$\mathcal{O}\left(n_\mathrm{ite}MN(\log_2N+\abs{\mathcal{L}})+MN(l_{\max}+1)^3\right)$}\\
        \midrule
        soft SIC-MMSE & {$\mathcal{O}\left(n_\mathrm{ite}MN\left(\log_2N+(l_{\max}+1)^3\right)\right)$}\\
        \bottomrule
    \end{tabular}
\end{table}

In practical channels, off-grid sampling of the matched filtered signal results in intrapulse spreading of DD domain symbols. While detection can be performed similarly to the on-grid case by considering the input-output relation of a sampled equivalent channel, the primary challenge lies in estimating the fractional delay and Doppler shifts. Specifically, because the DD domain channel takes effect in the form of twisted convolution, the channel response directly read off from the sampled response of an impulse pilot exhibits a phase error \cite{Tong2024ODDM_PhyChan}. Then, super-resolution channel estimation schemes are required, which may produce non-Gaussian channel estimation errors with inter-path correlation \cite{Khan2023ChanEst_Offgrid}, significantly complicating the performance degradation of iterative detectors. However, our analysis remains valid if the channel estimation errors can be approximated as Gaussian, following the central limit theorem.

\section{Conclusion}\label{sec:conclusion}
This paper provides a comprehensive comparison of three representative detectors for ODDM, namely MPA, MRC, and SIC-MMSE. By examining their detection accuracy, convergence behavior, and robustness against imperfect receiver CSI, we have offered insights into the trade-offs between complexity and BER performance for each detection algorithm. We present a detailed analysis of the post-equalization SINR and BER of MRC and SIC-MMSE, where channel estimation errors are considered. Our results show that both MPA and MRC suffer from error floors due to intractable residual interference. On the other hand, soft SIC-MMSE has the best BER performance, with hard SIC-MMSE following behind. In addition, MPA is more sensitive to channel estimation errors. Based on these results, we propose MRC-SD and SSMI-MRC to improve the BER performance of MRC. MRC-SD effectively lowers the error floor of MRC with a negligible increase in detection complexity. SSMI-MRC outperforms hard SIC-MMSE in BER while maintaining the same computational complexity. Our study extends the current understanding of the performance of ODDM detectors in practical system models.


%

\appendices

\section{Proof of Proposition \ref{proposition:sinr_mrc}}\label{appendix:psi_eta}

Recall the channel estimation error $\Delta\mathbf{g}_{q,\Delta l}$ and its variance $\sigma_{\Delta g}^2$ in (\ref{eq:ghat_vec}). The derivations of the terms in (\ref{eq:psi_mrc}) are as follows: The second term in (\ref{eq:psi_mrc}) is computed as
\begin{align}
    2\mathbf{g}_q^\herm\mathbf{g}_q\mathbb{E}\left[\Delta\mathbf{g}_q^\herm\Delta\mathbf{g}_q\right] = 2(l_{\max}+1)\sigma_{\Delta g}^2\mathbf{g}_q^\herm\mathbf{g}_q.
\end{align}
The third and fourth terms in (\ref{eq:psi_mrc}) are computed as
\begin{align}
    \mathbb{E}\left[\mathbf{g}_q^\herm\Delta\mathbf{g}_q\Delta\mathbf{g}_q^\herm\mathbf{g}_q\right] = \mathbb{E}\left[\Delta\mathbf{g}_q^\herm\mathbf{g}_q\mathbf{g}_q^\herm\Delta\mathbf{g}_q\right] = \sigma_{\Delta g}^2\mathbf{g}_q^\herm\mathbf{g}_q.
\end{align}
The fifth term in (\ref{eq:psi_mrc}) is computed as
\begin{align}
    \mathbb{E}\left[\bigl(\Delta\mathbf{g}_q^\herm\Delta\mathbf{g}_q\bigr)^2\right] &= \sum_{l\in\hat{\mathcal{L}}}\mathbb{E}\left[\abs{\Delta g_q[l]}^4\right] +\sum_{\mathclap{l_1,l_2\in\hat{\mathcal{L}},l_1\neq l_2}}\mathbb{E}^2\left[\abs{\Delta g_q[l]}^2\right]
    \nonumber\\
    &=\left(l_{\max}^2+3l_{\max}+2\right)\sigma_{\Delta g}^4.
    \label{eq:dgdg2}
\end{align}

The derivations of the terms in (\ref{eq:eta_mrc}) are as follows: By substituting (\ref{eq:z_hat}), the first term in (\ref{eq:eta_mrc}) is computed as
\begin{align}
        \mathbb{E}&\left[\mathbf{g}_q^\herm\hat{\mathbf{z}}_q^{(i)}\left(\hat{\mathbf{z}}_q^{(i)}\right)^\herm\mathbf{g}_q\right] = \mathbb{E}\left[\left(\mathbf{g}_q^\herm\hat{\mathbf{z}}_q^{(i)}\right)\left(\mathbf{g}_q^\herm \hat{\mathbf{z}}_q^{(i)}\right)^*\right]
        \nonumber\\
        &=\left(\sigma_e^2\right)^{(i)}\sum_{\mathclap{l_1,l_2\in\hat{\mathcal{L}}}}g_q^*[l_1]g_q[l_2]\sum_{\mathclap{\Delta l<0,\Delta l\in\dot{\mathcal{L}}}}g_{q,\Delta l}[l_1]g_{q,\Delta l}^*[l_2]
        \nonumber\\
        &\quad+\left(\sigma_e^2\right)^{(i-1)}\sum_{\mathclap{l_1,l_2\in\hat{\mathcal{L}}}}g_q^*[l_1]g_q[l_2]\sum_{\mathclap{\Delta l>0,\Delta l\in\dot{\mathcal{L}}}}g_{q,\Delta l}[l_1]g_{q,\Delta l}^*[l_2]
        \nonumber\\
        &\quad+\left(\sigma_e^2\right)^{(i)}\sum_{l\in\hat{\mathcal{L}}}\abs{g_q[l]}^2\sum_{\mathclap{\Delta l<0,\Delta l\in\dot{\mathcal{L}}}}\var(\Delta g_{q,\Delta l}[l])
        \nonumber\\
        &\quad+\left(\sigma_e^2\right)^{(i-1)}\sum_{l\in\hat{\mathcal{L}}}\abs{g_q[l]}^2\sum_{\mathclap{\Delta l>0,\Delta l\in\dot{\mathcal{L}}}}\var(\Delta g_{q,\Delta l}[l])
        \nonumber\\
        &\quad+P_t\sum_{l\in\hat{\mathcal{L}}}\abs{g_q[l]}^2\sum_{\mathclap{\Delta l\neq0,\Delta l\in\dot{\mathcal{L}}}}\var(\Delta g_{q,\Delta l}[l]) + \sigma_z^2\mathbf{g}_q^\herm\mathbf{g}_q.
\end{align}
By substituting (\ref{eq:z_hat}), the second term in (\ref{eq:eta_mrc}) is computed as
\begin{align}
    \mathbb{E}&\left[\Delta\mathbf{g}_q^\herm\hat{\mathbf{z}}^{(i)}_q\left(\hat{\mathbf{z}}^{(i)}_q\right)^\herm\Delta\mathbf{g}_q\right] = \mathbb{E}\left[\abs{\Delta\mathbf{g}_q^\herm\hat{\mathbf{z}}^{(i)}_q}^2\right]
    \nonumber\\
    &=(l_{\max}+1)\sigma_{\Delta g}^2\sigma_z^2
    \nonumber\\
    &\quad+\left(\sigma_e^2\right)^{(i)}\sum_{l\in\hat{\mathcal{L}}}\var(\Delta g_q[l])\sum_{\mathclap{\Delta l<0,\Delta l\in\dot{\mathcal{L}}}}\abs{g_{q,\Delta l}[l]}^2
    \nonumber\\
    &\quad+\left(\sigma_e^2\right)^{(i-1)}\sum_{l\in\hat{\mathcal{L}}}\var(\Delta g_q[l])\sum_{\mathclap{\Delta l>0,\Delta l\in\dot{\mathcal{L}}}}\abs{g_{q,\Delta l}[l]}^2
    \nonumber\\
    &\quad+\left(\sigma_e^2\right)^{(i)}\sum_{l\in\hat{\mathcal{L}}}\var(\Delta g_q[l])\sum_{\mathclap{\Delta l<0,\Delta l\in\dot{\mathcal{L}}}}\var(\Delta g_{q,\Delta l}[l])
    \nonumber\\
    &\quad+\left(\sigma_e^2\right)^{(i-1)}\sum_{l\in\hat{\mathcal{L}}}\var(\Delta g_q[l])\sum_{\mathclap{\Delta l>0,\Delta l\in\dot{\mathcal{L}}}}\var(\Delta g_{q,\Delta l}[l])
    \nonumber\\
    &\quad+P_t\sum_{l\in\hat{\mathcal{L}}}\var(\Delta g_q[l])\sum_{\mathclap{\Delta l\neq0,\Delta l\in\dot{\mathcal{L}}}}\var(\Delta g_{q,\Delta l}[l]).
\end{align}
The third term in (\ref{eq:eta_mrc}) is computed as
\begin{align}
    P_t\mathbb{E}\left[\mathbf{g}_q^\herm\Delta\mathbf{g}_q\Delta\mathbf{g}_q^\herm\mathbf{g}_q\right] = P_t\sigma_{\Delta g}^2\mathbf{g}_q^\herm\mathbf{g}_q.
\end{align}
Similar to (\ref{eq:dgdg2}), the fourth term in (\ref{eq:eta_mrc}) is computed as
\begin{align}
    P_t\mathbb{E}\left[(\Delta\mathbf{g}_q^\herm\Delta\mathbf{g}_q)^2\right] = P_t\left(l_{\max}^2+3l_{\max}+2\right)\sigma_{\Delta g}^4.
\end{align}





\ifCLASSOPTIONcaptionsoff
  \newpage
\fi



\bibliographystyle{IEEEtran}
\bibliography{references}
%

\end{document}